\documentclass[aps,twocolumn,nofootinbib,floatfix]{revtex4-2}
 
\usepackage[utf8]{inputenc}
\usepackage{graphicx}
\usepackage{xcolor}
\usepackage{url}
\usepackage{amsmath}
\usepackage{enumitem}
\usepackage[section]{placeins}  
\usepackage{bbm}
\usepackage{graphicx}
\usepackage{cancel}
\usepackage[utf8]{inputenc}
\usepackage[english]{babel}
\usepackage{tikz} 
\usepackage{dcolumn}  
\usepackage{mathtools} 
\usepackage{empheq}
\usepackage{bm}        
\usepackage{amssymb}    
\usepackage{bbold}
\usepackage{relsize}
\usepackage[T1]{fontenc}
\usepackage{mathtools}
\usepackage{amsfonts,amssymb,amsthm}
\usepackage{xcolor}

\DeclareMathOperator{\PP}{\cancel{P}}
\DeclareMathOperator{\CC}{\cancel{C}}

\def\cite#1{\citep{#1}}

\newcommand{\be}{\begin{equation}}
\newcommand{\ee}{\end{equation}}
\newcommand{\bea}{\begin{eqnarray}}
\newcommand{\eea}{\end{eqnarray}}
\newcommand{\eg}{{\it e.g.}}

\usepackage{ulem}
 
\begin{document}
 
\title{Mechanisms for spontaneous symmetry breaking in developing visual cortex}
  
\author{Francesco Fumarola}
\email{francesco.fumarola@riken.jp}
\affiliation{Laboratory for Neural Computation and Adaptation, RIKEN Center for Brain Science, 2-1 Hirosawa, Wako, Saitama 351-0198, Japan}
\author{Bettina Hein}
\affiliation{ Center for Theoretical Neuroscience, College of Physicians and Surgeons and Mortimer B. Zuckerman Mind Brain Behavior Institute, Columbia University, New York, NY}
\author{Kenneth D. Miller}
\email{kdm2103@columbia.edu}
\affiliation{Center for Theoretical Neuroscience, College of Physicians and Surgeons and Mortimer B. Zuckerman Mind Brain Behavior Institute, Columbia University, New York, NY}
\affiliation{Dept. of Neuroscience, Swartz Program in Theoretical Neuroscience, Kavli Institute for Brain Science, College of Physicians and Surgeons and Mortimer B. Zuckerman Mind Brain Behavior Institute, Columbia University, New York, NY}

\begin{abstract}
For the brain to recognize local orientations within images, neurons must spontaneously break the translation and rotation symmetry of their response functions -- an archetypal example of unsupervised learning. The dominant framework for unsupervised learning in biology is Hebb's principle, but how Hebbian learning could break such symmetries is a longstanding  biophysical riddle. Theoretical studies agree that this should require inputs to visual cortex to invert the relative magnitude of their correlations at long distances. Empirical measurements have searched in vain for such an inversion, and report the opposite to be true.
We formally approach the question through the hermitianization of a multi-layer model, which maps it into a problem of zero-temperature phase transitions. 
In the emerging phase diagram, both symmetries break spontaneously as long as (1) recurrent interactions are sufficiently long-range and (2) Hebbian competition is duly accounted for. The relevant mechanism for symmetry breaking is competition among connections sprouting from the same afferent cell. Such competition, and not the structure of the inputs, is capable of triggering the broken-symmetry phase required by image processing. We provide analytic predictions on the relative magnitudes of the relevant length-scales needed for this novel mechanism to occur. 
These results reconcile experimental observations to the Hebbian paradigm, shed light on a new mechanism for visual cortex development, and contribute to our growing understanding of the relationship between learning and symmetry breaking.
\end{abstract}
 
\date{\today}
\maketitle

\section{Introduction}

The primary visual cortex (V1) -- the first receiving area in the cerebral cortex for visual sensory information -- receives signals from the lateral geniculate nucleus of the thalamus (LGN), which in turn receives signals directly from the eyes. Both LGN and V1 extend in two dimensions so as to embody a continuous map of the world as seen through the two eyes. In other words, these brain regions are arranged ”retinotopically”, each neuron responding to input in the vicinity of a certain point on the retina, with neighboring areas of the retina represented by neighboring neural areas. Moreover, cells in LGN can be excited either by light onset or by light offset on the corresponding spot of the retina, and cells of these two types are known respectively as ON-center and OFF-center cells. 
 
We will think of the instantaneous visual stimulus as pixel values for each position in the two-dimensional retinotopic space. The ”receptive field” (RF) of a visual neuron is defined as the linear kernel that best determines its activity as a function of the visual stimulus. LGN cells, like retinal cells~\citep{Kuffler,Barlow}, have RFs that are roughly circularly symmetric~\citep{HubelWiesel1}, while V1 neurons best respond to a particular orientation of a light/dark edge~\cite{HubelWiesel2,Hubel_Wiesel62}. Furthermore, in response to a drifting periodic luminance grating, the temporal mean of the total LGN input to a V1 cell is untuned for orientation, while the size of the temporally periodic modulation about the mean of this input is orientation tuned~\citep{Ferster_Miller00,Lien_Scanziani13}, which indicates that deviations from circular symmetry in the RFs of individual LGN cells do not contribute appreciably to the orientation selectivity of V1 cells \citep{Troyer_etal98}. Instead, rotational symmetry  is broken by the spatial arrangement of the set of LGN cells that make synaptic connections onto a given V1 neuron. As originally postulated by \citet{Hubel_Wiesel62}, this spatial arrangement appears to involve spatially adjacent subregions alternating between ON-center and OFF-center inputs \citep{Reid_Alonso95,Ferster_Miller00}, representing adjacent retinotopic subregions in which light or dark stimuli, respectively, best drive the V1 cell. Here we focus on understanding the origin of this spatial arrangement that breaks rotational symmetry.

The orientation preferences of V1 cells are locally continuous, and rotate roughly periodically with movement along the two dimensions representing retinotopy, in all non-rodent species that have been studied [Ref. \citealp{HubelWiesel3}; reviewed in Ref. \citealp{Kaschube_etal10}]. This arrangement of orientations over cortical space is known as an "orientation map". We will also address the breaking of translational symmetry that is necessary for such a map.

The sensitivity to local orientation develops in V1 cells without visual experience~\cite{HubelWiesel3,Huberman_etal08} but depends on normal patterns of spontaneous activity (activity without vision, \eg\ in the dark) in LGN and V1~\citep{Huberman_etal08}. For this and other reasons, V1 orientation selectivity is thought to arise from a process of activity-dependent self-organization, most likely instructed by the spontaneous activity patterns~\citep{Miller_etal99,Kaschube_etal10}. The problem is thus placed within the general framework of symmetry breaking during learning, a branch of theoretical physics that recently achieved substantial progress~\cite{fok2017spontaneous,wu2020phase,gordon2020relevance,tanaka2021noether,kline2021gaussian}.

Understanding the early LGN-V1 dynamics is a renewed concern at the moment because the engineering of artificial neural networks has made progress through an increasingly detailed mimicking of visual cortex~\cite{borji2020shape,evans2021biological,dapello2020simulating}. Such progress has been buttressed by an ever closer mapping between brain vision and computer vision notably insofar as it concerns unsupervised learning~\cite{zhuang2021unsupervised,nayebi2021unsupervised,rolls2021learning,xiong2020loco}.

On one hand, a first stage of unsupervised learning is often used as a regularizing technique for weights initialization~\cite{paine2014analysis,erhan2010does}, mimicking the lack of annotations in the early infancy of animals; on the other hand, the reliance on LGN's spontaneous activity has been matched by the machine-learning practice of pre-training a network on more generic data before one begins to train it on the data of interest~\citep{mahajan2018exploring,kataoka2020pre}.
Engineering practices like those are likely indeed to serve similar efficiency goals in visual cortex; thanks to them, newborn animals start learning to see even before their eyes open.

How unsupervised learning is implemented in the nervous system is a question that dates back to the 19th century, when researchers from various areas converged to the notion that associating information was the quintessential brain function~\cite{alexander1855senses,ribot1870,james1890principles}. The first concrete biophysical paradigm for unsupervised learning was proposed 70 years ago, independently by Hebb~\cite{hebb} and Konorski~\cite{k}. Hebbian learning, as it became known, posits that synapses are strengthened by temporal correlation between their pre- and postsynaptic patterns of activity ("cells that fire together wire together")~\cite{shatz1992developing,lamprecht2004structural,markram2011history}. It plays in brain science a similar role to Principal Component Analysis (PCA) in data science and, indeed, PCA and Hebbian learning are largely equivalent formulations~\cite{principe1999neural}. 

The development of orientation selectivity was one of the earliest testing grounds for Hebbian learning~\cite{vdMalsburg,Linsker86,Miller_etal86}. Given that orientation selectivity develops normally without vision, it is the correlations in spontaneous activity that must be taken into account by this type of theory. Early Hebbian models of V1 used either a specific set of inputs~\cite{vdMalsburg} or an ensemble average over inputs~\cite{Linsker86,Miller_etal86,Miller_etal89a,miller94model} to simulate the behavior of cells receiving feedforward input through Hebbian synapses from LGN. The model of~\cite{Linsker86} was analyzed in more detail by~\cite{MacKayMiller1}, who studied numerically the dominant eigenmodes of the time-evolution operator, besides proving general theorems on
their properties. In~\cite{miller94model}, it was shown that the functional organization of orientation preferences across the cortex could arise through Hebbian dynamics from a so-called Mexican-hat profile of input correlations, in which same-center-type inputs are more correlated than opposite-center-type at short separations and the converse is true at longer separations.

The rationale for this assumption comes from the general
fact that the RFs of ON cells consist of a circular, ON (light-preferring) center and an OFF (dark-preferring) surround that forms a ring about the center\cite{Kuffler,Barlow,HubelWiesel1}, and similarly OFF cells have an OFF center and an ON surround. The correlations between two LGN cells will therefore be null if they are located at such a large distance that their RFs do not overlap; at distances where the surround of one cell overlaps with the center of the other one, correlations would be expected to be negative for same-type cells, positive otherwise; finally, for distances short enough as to involve a substantial overlap of the two centers, correlations should be positive for same-type cells, negative otherwise (see~\cite{miller94model}, Fig. 3). This scenario leads naturally to a zero-crossing trend in the correlation functions. 
Two inputs of the same center-type (ON/ON or OFF/OFF) will be more positively correlated than two opposite-type inputs (ON/OFF) at short retinotopic separations, but less positively correlated at larger retinotopic separations. Thus, the difference between the same-type and opposite-type correlation functions should have a Mexican hat shape. 
    
With this premise, the model 
of~\cite{miller94model} predicts that orientation selectivity will arise via activity-dependent competition between ON- and OFF-center inputs. Hebbian plasticity leads an individual cell to receive a well-correlated set of inputs. This yields a set of inputs to each cell that alternates between inputs of one center-type and the other, with a spatial period corresponding to the alternation between same-type and opposite-type pairs being best correlated. In addition, local excitatory connections between cortical cells lead nearby cells to develop similar preferred orientation, eliminating high spatial frequencies in the map of orientations.  At larger distances different orientations arise from the random initial conditions, so that a low-pass map arises. The actual maps are bandpass (periodic) rather than low-pass \citep{Kaschube_etal10}. Thus, the model does not produce realistic map structure but does show the breaking of translational symmetry and elimination of high spatial frequencies seen in real maps.
 
Theories of Hebbian development similar to the one explored in \cite{miller94model} were studied by \cite{Miyashita,Yamazaki,Wimbauer}. In \cite{Wimbauer}, a rigorous proof  was worked out that, indeed, ”in order to get orientation-selective receptive fields, the spatial correlation function of the inputs that drive the development must have a zero crossing.” 
 
This scenario, however, has 
been called into question by direct measurements of correlations in LGN activity as a function of retinotopic distance \cite{OhshiroWeliki}. Experiments on young ferrets, at the ages over which orientation selectivity develops, found no
Mexican hat in LGN. Instead, same-type pairs were more strongly correlated then opposite-type pairs at all retinotopic separations, and the decay with distance was monotonic. 

Measurements were taken at various developmental times and in the presence of various
visual inputs. The only hint of the existence of a zero-crossing in the correlation
functions appears with white pixelated noise as visual input (not in spontaneous activity absent a visual stimulus), and then only in more mature animals in which orientation maps have already developed. These measurements have posed an essential problem for the current understanding of learning in the brain,  the resulting question being whether visual cortex can develop at all through Hebb's principle, which motivated the present study.

\section{Results} 
\label{sec:results}

A fundamental invariance law for models of brain vision is that a  simultaneous translation/rotation of the animal and of the image it views will not affect brain activity~\citep{Bressloff_etal01}. More specifically, a simultaneous translation/rotation of primary visual cortex and of its input source should not affect the resulting RFs. These two symmetries (translation and rotation) lead to three relevant symmetry classes of the solution, detailed in table~\ref{table:phases}. It remains to be understood which phases are allowed by biological mechanisms. 

\begin{table}[htp] 
\caption{Symmetry classes of the solution} 
\centering      
\begin{tabular}{c|c|c|}
\hline\hline                        
Phase Label & Receptive Field & Orientation Selectivity \\ 
\hline  
N &  uniform across cortex & non-selective \\ 
\hline   
R &  uniform across cortex & selective \\ 
\hline     
T  & varies across cortex & selective  \\
\hline\hline
\end{tabular} 
\label{table:phases}
\end{table} 

\begin{figure}[htp]
\includegraphics[width=0.45\textwidth]{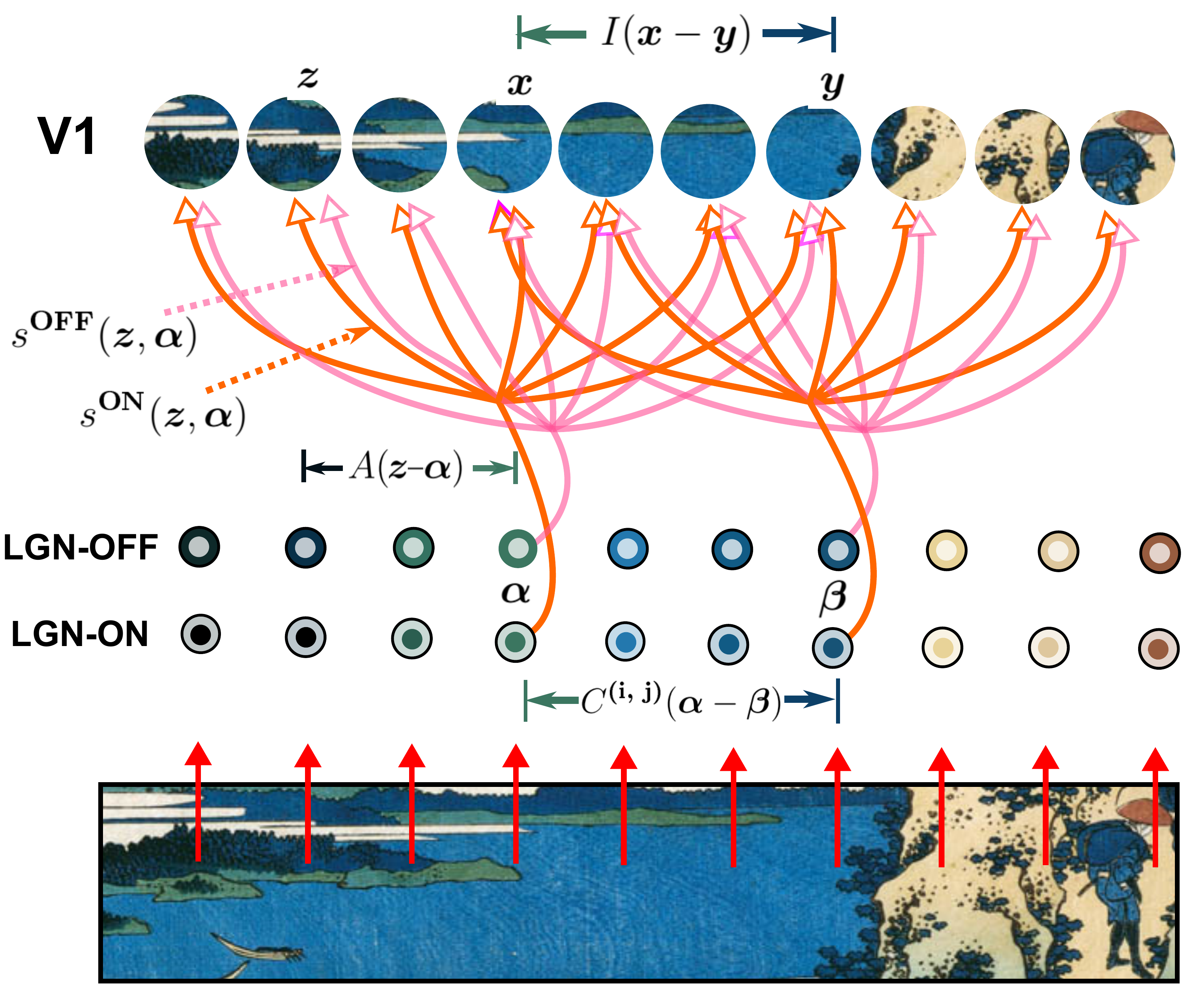}
\centering 
\caption{\small{\textbf{Schematic depiction of the model.}} 
Primary visual cortex (V1) is  the first receiving area in the cerebral cortex for visual sensory information. Here it is represented by a single post-synaptic layer of cells, depicted by the upper row of larger circles. Inputs to V1 come from the lateral geniculate nucleus of the thalamus (LGN), which in turns takes input from the eyes. LGN cells are excited either by light onset or by light offset on the corresponding spot of the retina, and these two types of cells are represented by the two rows of smaller circles. The quantity of synapses from an LGN location $\bm{\alpha}$ to a V1 location $\bm{z}$ is described by an "arbor density" $A(\bm{z}-\bm{\alpha})$. Their total synaptic strength is defined as either of the functions
$s^{\text{ON}}(\bm{z},\bm{\alpha})$ or $s^{\text{OFF}}(\bm{z},\bm{\alpha})$ depending on whether the presynaptic cell is of the ON or OFF type. 
Correlations between the activities of two presynaptic cells  of types $\text{(i)} \in \{  \text{ON, OFF}\}$ and  $\text{(j)} \in \{ \text{ON, OFF}\}$ located at retinotopic positions 
$\bm{\alpha}$ and $\bm{\beta}$ are described by a set of functions $C^{\text{(i,j)}}(\bm{\alpha} - \bm{\beta})$, while the effect, via lateral connections, of activity at cortical position $\bm{x}$ on activity at cortical position $\bm{y}$, is characterized by the function $I(\bm{x}- \bm{y})$.}
\label{fig:model_cartoon}
\end{figure}

We address this question with the model illustrated in Fig.~\ref{fig:model_cartoon}, in which V1 is represented by a single post-synaptic layer, and the lateral geniculate nucleus (LGN) by two overlapping layers containing ON-center and OFF-center cells respectively. The layers are modeled as infinite planes, endowed with a retinotopic metric and inhabited by a continuum of cells.

Synapses from a given afferent cell of either type are said to belong to an "arbor"~\cite{Jacobson} and the number of them targeting a given cortical cell is  a function $A(\bm{r})$ of the retinotopic distance vector $\bm{r}$ between the two cells. For definiteness we take $A(\bm{r})$ to be Gaussian, decaying over an "arbor width " $\rho$. 

Correlations between the activities of two presynaptic cells  of types $(i)=  \text{ON/OFF}$ and  $(j)= \text{ON/OFF}$ are described by a function $C^{(i,j)}(\bm{r})$ of the retinotopic distance vector $\bm{r}$. 
This function must be  monotonically decreasing by the experimental results discussed in the Introduction, and may be thus assumed to be a Gaussian with a given correlation length  $\eta$.  A similarly Gaussian ansatzes is adopted for the strength for lateral connections between cortical neurons, characterized through an interaction length-scale $\zeta$ (see Appendix~\ref{sec:model_setup}). The phase diagram can thus be plotted using for coordinates any pair of dimensionless ratios among the  three parameters $\rho,\eta,\zeta$. 

Let $s^{\text{ON}}$ and $s^{\text{OFF}}$ be the synaptic strength of connections from ON and OFF cells to cortex, as per Fig.~\ref{fig:model_cartoon}. Taking  synaptic plasticity to be slow on the time scale of neural activity, we  obtain independent equations for their sum $s^{(S)}\equiv s^{\text{ON}}+s^{\text{OFF}}$ and difference $s^{(D)}\equiv s^{\text{ON}}-s^{\text{OFF}}$ (see Appendix~\ref{sec:model_setup}). Our interest is in the development of orientation selectivity via the formation of alternating RF subregions in which ON or OFF LGN inputs, respectively, are dominant. Hence we are interested in development of a pattern in $s^{(D)}$, while $s^{(S)}$ is not expected to form interesting structure. Thus, we focus on computing for $s \equiv s^{(D)}$.

We would like to incorporate the fact that correlation-based development is competitive, meaning that when some synapses grow stronger, others grow weaker. Beginning with the first model of activity-dependent development of \citet{vdMalsburg}, theorists have often modeled this competition as a constraint that the total synaptic weight received by a neuron is conserved. Alternatively, it can be modeled as a homeostatic process that maintains the average postsynaptic firing rate of neurons about some set point, as is seen experimentally \citep{turrigiano2011too}. 
Such a constraint upon the summed postsynaptic strength will alter the equation for the sum $s^{(S)}$, but not for the difference $s^{(D)}$. Since we are concerned with the development of $s^{(D)}$, we will ignore such a constraint here.

Another form of competition is that presynaptic axonal arbors compete for postsynaptic connections. In many neural systems, it has been shown that if an arbor loses overall synaptic strength, it competes more effectively to retain it, while if it has too much, it competes less effectively, so that arbor retraction takes place (for reviews, see \cite{LuoOLeary} and \cite{Jacobson}, chapter 8). While the physics of competition among LGN arbors for cortical innervation is still unclear, it seems reasonable to assume that, given statistically equal activity of LGN cells, it cannot be the case that some arbors lose most of their innervation to cortex while others take over. Rather, all arbors should retain roughly equal innervation. This can also be thought of as a homeostatic process, maintaining the overall projection strength of each neuron about some set point.

Since axonal competition will separately constrain the overall strength of ON innervation and of OFF innervation, it will constrain the development of $s^{(D)}$. We model this as a precise conservation of the total strength of synapses projected by each arbor (see Appendix~\ref{sec:model_setup}). The introduction of homeostatic constraints, which is demanded by biophysical considerations,  considerably complicates the problem. Indeed, such constrained models have been predominantly an object of numerical investigations (for the case of a single V1 cell see also~\citep{Miller_MacKay94}), and these happened to miss their most relevant potentiality -- the emergence of a fully symmetry-breaking and hence orientation-selective and map-forming phase without `Mexican-hat'-like functions (which drive development of periodic patterns) in either correlations or intracortical interactions.
 
\begin{figure}
\includegraphics[width=0.45\textwidth]{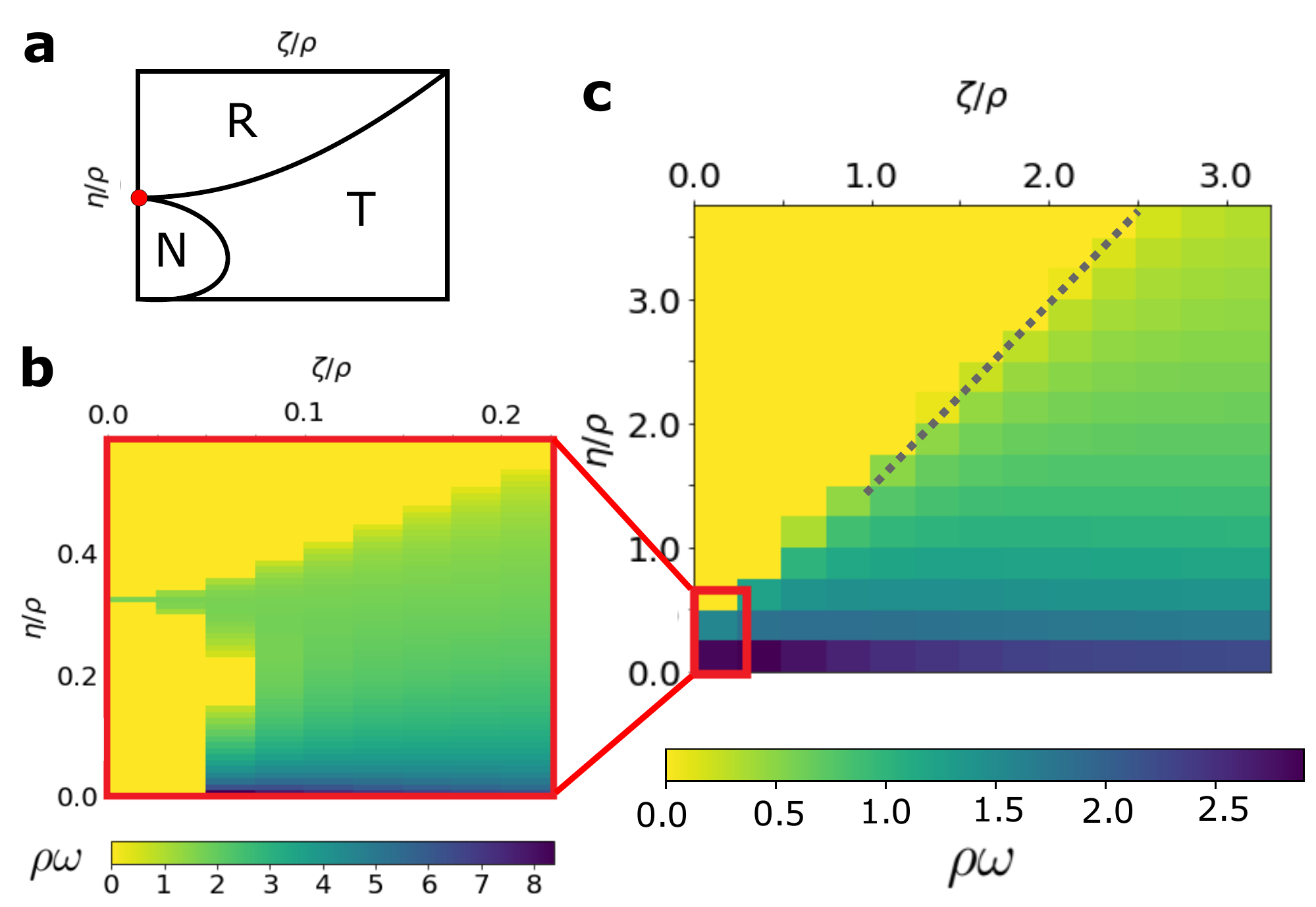}
\centering
\caption{
\small{\textbf{Phase diagram of the model.}} \textbf{a}, Structure of the phase diagram resulting from analytical investigation of the time evolution. The model's parameters are the correlation length $\zeta$ of the inputs from LGN, the radius of outgoing connections $\rho$, and the length-scale $\eta$ of lateral connections in V1. A phase diagram may use for coordinates any pair of dimensionless ratios among these parameters; here $\zeta/\rho$ and $\eta/\rho$ are used. The labels N,R,T correspond respectively to no symmetry breaking, only-rotational symmetry breaking, and rotation-translation symmetry breaking (cf. Table~\ref{table:phases}).
The red dot marks the inferred location of a triple point.  
\textbf{b}, Dominant wavenumber of map modulation across cortex in units of the inverse arbor radius. 
The preservation of cortically uniform states is confirmed in the regions corresponding to the R and N phases. Clearly visible are the sharp boundaries of the two cortically uniform phases N and R, separated by the non-uniform phase T, where orientation maps arise.  
The plots were obtained by diagonalizing a discretized version of the operator $L^p$ defined by Eq.~(\ref{main}), where RFs were confined to a square of side equal to $6$ arbor radii represented by a $15\times 15$ grid. 
\textbf{c}, Larger-scale view of the phase diagram, displaying as a gray dotted line  the asymptotic phase boundary $\eta_c \sim \sqrt{2} \zeta$ obtained from analytic solution of the model.   
}
\label{fig:phaseDiagram}
\end{figure}

As detailed in Appendices~\ref{sec:homeostatic}-\ref{sec:projection_operators}, after incorporating the above constraints the time-evolution operator can be hermitianized and thus mapped into a quantum-mechanical Hamiltonian whose low-energy states  describe the long-term relaxation of the system.
Quantum-mechanical tools allow then to see that  all three phases listed in table~\ref{table:phases} enter the phase diagram, whose  structure is illustrated by Fig.~\ref{fig:phaseDiagram}, panel A. 

\begin{figure}
\includegraphics[width=\textwidth]{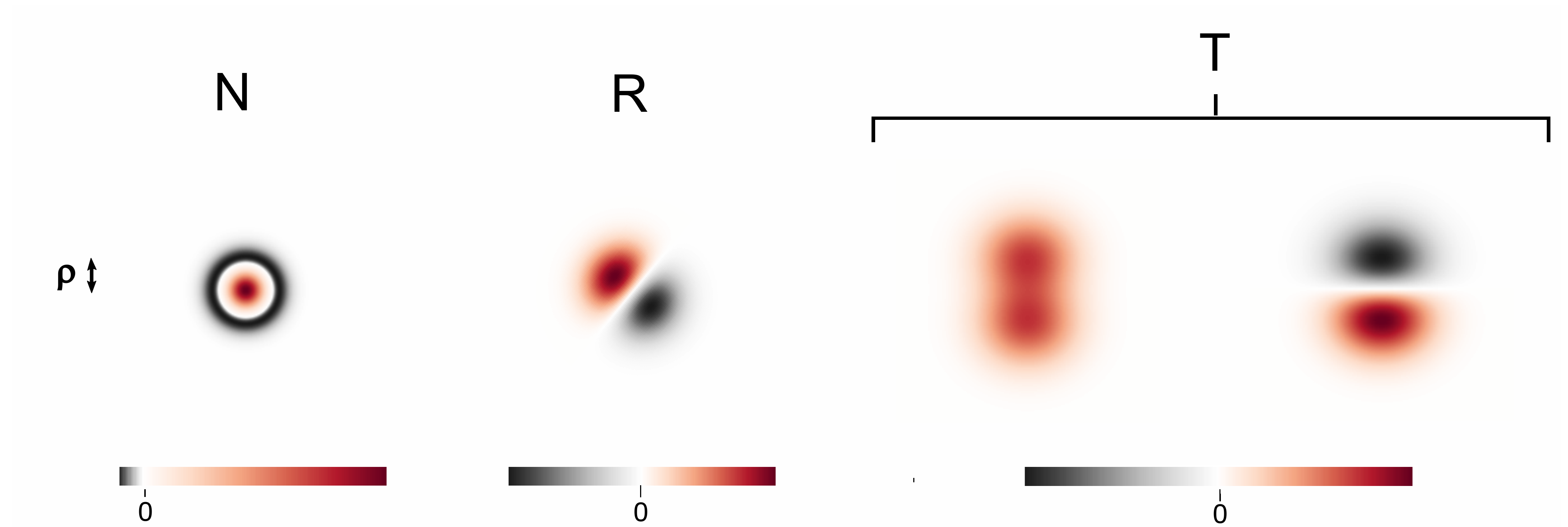}
\centering 
\caption{\small{\textbf{Receptive fields by symmetry class.}} Plots of the receptive fields of  formulas~(\ref{psiN}),~(\ref{psiR}), and~(\ref{psiT}) respectively for the N-, R- and T-phase. Hue scale ranges between minimum and maximum values, allowing for an arbitrary scaling factor. Functions were evaluated at representative points $P = (\zeta/\rho,\eta/\rho)$ in parameter space given by  $P_N=(.02,.2)$, $P_R=(.05,.7)$, $P_T=(5,3)$ (see Supplemental Material for numerical results with the same parameters). The relative length of an arbor radius $\rho$ for all plots is shown to the left of the figure. The receptive field for the R-phase is plotted with a randomly chosen orientation. For the T-phase, the function $s_{\text{T}}$ of Eq.~(\ref{psiT}) was rotated by the complex angle $\phi_0 =\arctan\left( - \int \Im s_{\text{T}} /\int \Re s_{\text{T}} \right)$ so as to make the imaginary part odd under inversion of the cortical modulation axis (the real part becomes symmetric as a consequence, see Supplemental Material Sec.~\ref{sec:long-term-dyn}) and we separately display its real (left) and imaginary (right) components. The fastest-growing mode of the time evolution is obtained multiplying this by a modulating phase factor $e^{i \omega x}$, with $x$ being the coordinate for the degenerate direction of modulation in cortical space.
}
\label{fig:real_imag_eigfs}
\end{figure}

For low values of both $\zeta/\rho$ and $\eta/\rho$ (in a lunette extending from the lowest stretch of the $\eta$-axis) cortex is in a symmetry-preserving state that we term "N-phase",  where the RF is identical and unoriented at all points in retinotopic space. This would mean that, upon eye opening, a given cell's responses could indicate the location of an object in visual space but could not indicate its orientation. 
In terms of the cortical position $\bm{x}$ and of the difference between pre- and post-synaptic positions $ \bm{r} $, a variational approximation to the RF in this phase is given by
\be  
\label{psiN}
s_N(\bm{x},\bm{r}) \propto  (R^2 - \bm{r}^2) \exp\left(- \frac{\bm{r}^2}{2 \hat{\rho}^2} \right)
\ee 
where $\hat{\rho} = \sqrt{2} \rho \left( 1 + \sqrt{1 + \frac{4 \rho^2}{\eta^2 +\zeta^2}}\right)^{-1/2} $ and $R \sim  \left(  \frac{4 + \sqrt{10}}{3} \right)^{1/2} \rho^{1/2} (\eta^2 + \zeta^2)^{1/4} $ (see Fig.~\ref{fig:real_imag_eigfs}).

For values of $\eta$ above a critical value which increases as a function of $\zeta$, cortex is in a state  where rotation symmetry is spontaneously broken at every point ("R-phase"). Since by contrast translation symmetry is not broken, the  RFs (including their orientation preferences) are the same everywhere. To an animal with this visual cortex, a pencil slanting at the right angle would be perceivable as such; even shifting it in front of its eyes would pose no hindrance to vision. But a major handicap would emerge if the pencil tilted at a different angle, as all cortical cells would be poorly responsive or unresponsive to this stimulus.

The receptive field in this phase has a degeneracy of order two which can be represented with the exact basis (see Fig.~\ref{fig:real_imag_eigfs})

\begin{equation}
\label{psiR}
s^{x,y}_R(\bm{x},\bm{r}) \propto \left\{ \begin{array}{c} 
r_x \\ r_y\end{array} \right\} \exp \left( - \frac{r^2}{ 2 \hat{\rho}^2} \right) 
\end{equation} 

The qualitatively novel phase reported in this paper ("T-phase") appears for sufficiently large values of $\zeta$. In the T-phase, a double spontaneous breaking of rotation and translation symmetries allows the animal to perceive in principle both shifts and rotations of an elongated object. This biologically plausible phase had previously been found only in the presence of finely tailored choices of the input's correlation function Eq.~\ref{gaussian_choice_corr}, such as differences between Gaussians of different widths~\cite{miller94model,Wimbauer} or similar \citep{Linsker86}, which the experiments discussed previously~\citep{OhshiroWeliki} argue against.

A dipole approximation on the relevant time-evolution operator, detailed in the Supplemental Material, reveals that the main curve partitioning the diagram (RT boundary) is asymptotically linear far from the origin ($\max(\zeta,\eta) \gg \rho$), and approximated by $\eta_c \sim \sqrt{2} \zeta$ (gray dotted line in~\ref{fig:phaseDiagram}B). For interaction lengths above that boundary the system breaks only rotation symmetry; beneath it,  phenomenologically relevant orientation maps emerge. 

On the T-side of the transition line, the RF is approximated by a linear combination of the real and imaginary parts of the function
\begin{eqnarray}
\label{psiT}
\hspace{-20pt} s_T(\bm{x}, \bm{r}) &\propto&  e^{- \frac{r^2}{2 \rho^2} - i \frac{\eta^2}{\mu^2} \bm{\omega} \bm{r} + i \bm{\omega} \bm{x}} 
\times \\  \nonumber && \hspace{20pt}
\times 
\left( 1 - e^{ - \frac{1}{2} \frac{\zeta^4}{\mu^4} \rho^2 \omega^2 - i \frac{\zeta^2}{\mu^2}  \bm{\omega} \bm{r}}  \right) 
\end{eqnarray}
where  $\mu=\sqrt{\eta^2+\zeta^2}$ and $\omega \equiv | \bm{\omega} |  \sim \frac{\mu}{\zeta^2} \sqrt{\frac{ 2 \zeta^2}{\eta^2} -1 } $ in the vicinity of the phase boundary.

The long-term behavior described by formula~\ref{psiT} (displayed in Fig.~\ref{fig:real_imag_eigfs})
is degenerate in the direction of the $\bm{\omega}$ vector, and will be summed over directions made available by the initial condition. Notice that, because $\hat{\rho} \sim  \rho$ for  large $\mu/\rho$, at the phase boundary the T-phase functions (\ref{psiT}) transform continuously into the projection of the R-phase eigenfunction (\ref{psiR}) on the axis parallel to $\bm{\omega}$. 

\onecolumngrid

\twocolumngrid
\onecolumngrid

\begin{figure}
\includegraphics[width=\textwidth]{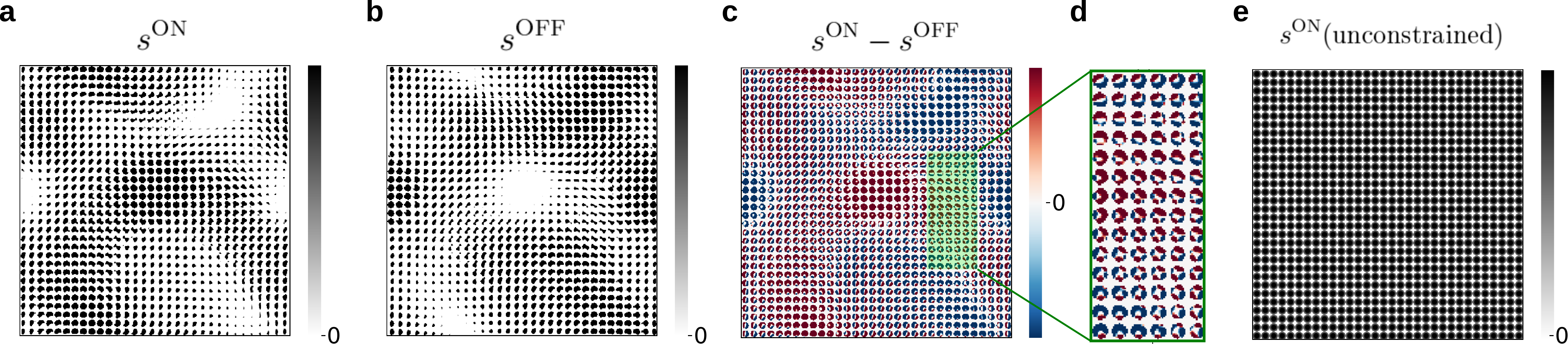}
\centering
\caption{\small{\textbf{Results of dynamical simulations.}} Simulations of the equations of motion (Eqs.~(\ref{sdequation}) of Methods) were performed on  discretized receptive fields, evolving from random initial conditions while waiting for the configurations to stabilize. We used $32\times 32$ grids of cortical cells and of ON and OFF LGN cells, with periodic boundary conditions. Each LGN cell projected an arbor of inputs to cortex that was nonzero over a circle of diameter 13 centered on its corresponding point in the cortical grid.
\textbf{a},~Large-times configuration of $s^{\text{ON}}$ after evolving  the constraint coming from Hebbian competition. Each 13 $\times$ 13 square contains the diameter-13 (or smaller) set of ON weights to one cortical cell. A 32 $\times$ 32 grid of cortical cells is illustrated. \textbf{b},~As in panel~\textbf{a}, but for $s^{\text{OFF}}$, which yields analogous breaking of the symmetries.
\textbf{c},~Difference between $s^{\text{ON}}$ and $s^{\text{OFF}}$ of panels~\textbf{a} and~\textbf{b}, with a portion expanded to highlight cell-by-cell structure (panel~\textbf{d}). \textbf{e},~Large-times state of $s^{\text{ON}}$ evolved \textit{without} the competitive constraint (Eq.~\ref{unconstrained_dyn} of Methods) and displaying indeed no trace of either symmetry breaking. In this fully symmetric solution, all RFs are identical and lacking orientation tuning. 
Parameters used for all these simulations: $\eta/\rho=0.75, \zeta/\rho=0.25, \rho=6.5$.}
\label{fig:or_maps}
\end{figure} 

\newpage
\twocolumngrid 

The point where this phase boundary hits the $\zeta=0$ axis is a triple point at which all three phases coexist. 
 
Finally, the above results have been confirmed by numerical investigation. Fig.~\ref{fig:phaseDiagram}B shows results for the dominant wavenumber of map modulation across cortex and confirms  the preservation of cortically uniform states in the regions corresponding to the R and N phases. The different symmetries of the RFs for the two cortically uniform regions  (predicted as in Figs.~\ref{fig:phaseDiagram}A) are readily confirmed by inspection (see Figs.~\ref{fig:N_check},\ref{fig:R_check},\ref{fig:T_check}).

We further performed numerical simulations of the equations of motion (Eqs.~(\ref{unconstrained_dyn})  both in the absence of any constraint and while imposing  the constraint (Eq.~\ref{conservation}). Simulations were carried out by evolving random initial conditions and waiting for the configurations to stabilize. The unconstrained case, displayed in Fig.~\ref{fig:or_maps}E, leads to a fully symmetric solution, with all RFs identical and lacking orientation tuning. For parameters picked in the T-phase, dynamical simulations in the presence of the constraint  yield symmetry breaking in both $s^{\text{ON}}$ (Fig.~\ref{fig:or_maps}A) and $s^{\text{OFF}}$  (Fig.~\ref{fig:or_maps}B) and ultimately in the quantity of interest, their difference (Fig.~\ref{fig:or_maps}C-D).

\section{Discussion}

Our understanding of  unsupervised learning in biological systems relies crucially on Hebb's principle - the statement that co-active cells strengthen mutual connections - and visual processing is a classical testing ground for that hypothesis. But the foundation of existing theories on the Hebbian development of the quintessential feature of visual cortex, orientation selectivity, is a sign-switch in the input correlations at large distances, of which experimental studies find no trace~\cite{OhshiroWeliki} . 

This motivated us to re-examine the problem with a focus on the role of competition among LGN inputs to cortex. We built and solved analytically a  model of receptive field development in which 
the total projection strength from each ON or OFF cell in LGN is constrained to remain constant. We showed that this constrained dynamics is able to produce orientation-selective RFs that vary smoothly across the cortex, even with  data-compatible input correlations that decay monotonically with distance, without a sign change.

The multilayer model we employed consists of two bottom layers representing ON/OFF cells in LGN and a top layer representing V1. The key dimensionless parameters are the widths of LGN correlations and of cortical lateral interaction, in units of the arbor radius. As the model is translationally and rotationally invariant, possible solutions can break either symmetry, or both, or none. Orientation selectivity requires breaking rotation symmetry, while variation of preferred orientations across cortex requires also the breaking of translation symmetry. We ignore the fact that intracortical connections develop in a pattern that breaks rotational symmetry \citep{Bosking_etal97,Schmidt_etal97b}, which yields other symmetry classes \citep{Bressloff_etal01}.

The uniformly non-selective phase (where neither symmetry is broken) prevails for sufficiently small values of the width parameters. Rotation invariance is broken for sufficiently long-range cortical interactions; translation invariance, for sufficiently long-range LGN correlations. Besides calculating the phase boundaries, we estimated the functional form of receptive fields in the various regimes and, for the T phase, which has nonzero cortical wavenumber, we evaluated explicitly the preferred value of this wavenumber across parameter space. 

Why does orientation selectivity develop without a sign change in correlations? Previous analyses~\citep{miller94model} focused on maximizing correlations within one RF, and implicitly considered interactions between RFs as a perturbation that coordinates the developing orientations and spatial phases between cells. For a maximally correlated RF to be oriented, the sign change is required. However, when cortical interactions are sufficiently long range, interactions between RFs can become equal to or dominant over interactions within RFs. Our results show that, in this case, between-RF correlations are maximized, given the constraint, by segregating ON and OFF subregions within RFs so that RF pairs at many distances can achieve overlaps of same-type subregions. This is energetically favored to the alternative which is to have a periodic alternation across cortex of all-ON and all-OFF RFs (with the period as large as possible while satisfying the constraint, as discussed for competition between inputs from the two eyes in~\citep{Miller_etal89a}).

The relevance of the study to the current understanding of cortical development lies in the demonstration that orientation selectivity and its smooth variation across cortex can develop without zero-crossing correlations. The role of competition between ON- and OFF-center LGN inputs  proves to be key, as opposed to previous theories where the structure of LGN activity rather than the homeostatic constraints would be responsible for spontaneous symmetry breaking in V1. 
This extends the applicability of Hebbian development, bridges it with advances in our experimental knowledge of V1, and casts light on the function of competition among axonal arbors for innervation, the biophysical mechanisms of which are still far from being properly understood~\cite{keck2017integrating}.

Our model shares with previous models the non-biological feature that the developed orientation maps are low-pass rather than bandpass (periodic). This is because lateral interactions that maximize correlation between RFs of cells that excite one another, and anti-correlation when one cell inhibits another, lead to periodic changes in the spatial phase of RFs rather than in their preferred orientation~\cite{miller94model}. These periodic phase changes maximize the spatial overlap of subregions of the same ON or OFF type to maximize correlations, and of opposite types to maximize anti-correlations. We have argued~\citep{Kayser_Miller02} that, if orientation selectivity develops in phase-selective cells (simple cells), periodic maps might develop through interactions between phase-nonselective cells (complex cells) receiving input from simple cells of multiple preferred phases; the influence of the complex cells would propagate back to the phase-selective cells to organize maps even as the phase-selective cells are responsible for the development of orientation selectivity itself. \citet{Antolik_Bednar11}~demonstrated such a scenario, but using some non-biological assumptions such as no interactions among developing simple cells. 
 
An extant direction for future studies will lie in the possible nonlinear complications of Hebbian models of cortical development, including simultaneous development of intracortical and input synapses. Nonlinear variants of Hebbian dynamics may be key to creating bottom-up models of orientation development that are able to reproduce  band-pass, periodic orientation maps like those measured in  cortex~\citep[\eg][]{Kaschube_etal02}. This may further serve to bridge Hebbian theory to Landau-type models of universal behavior such as those of~\cite{Kaschube_etal10}.

By addressing the riddle of orientation selectivity in biological networks, our work suggests some possible new avenues for research on artificial networks, which have been seen to perform unsupervised learning in a strikingly brain-like fashion~\cite{zhuang2021unsupervised}. 
Future research in this sense may belong to two directions: 
 (1) a comparison of the multilayer dynamics we demonstrated to the learning trajectories of units in convolutional networks notably during pre-training (along the lines of~\cite{yamins2014performance,lindsay2017understanding,schrimpf2020brain}) to  shed light on the degree of universality of the mechanisms discussed; (2) the deployment of the competitive constraints we demonstrated toward engineering convolutional networks for unsupervised learning such as those of~\cite{miclut2014committees,dundar2015convolutional,knyazev2017recursive,zbontar2021barlow,chen2020simple}, in analogy with performance-boosting via cortex-like bottlenecks in~\cite{dapello2020simulating}.

\section{Credits and Acknowledgments} 

F.F. worked on the analytics, B.H. on the numerics, and K.D.M. supervised the project. All authors have reviewed the manuscript. We benefited from conversations with Tadashi Yamazaki,  Tomokazu Ohshiro, and Hideaki Shimazaki. This research has been supported by NSF (award DBI-1707398) and by the Gatsby Charitable Foundation (GAT3708).  FF has also been supported by RIKEN Center for Brain Science, Brain/MINDS from Japan's Agency for Medical Research and Development under Grant No. JP20dm020700, and JSPS KAKENHI Grant No. JP18H05432.

\vspace{30pt}
\centerline{\textbf{APPENDIX}}
\setcounter{equation}{0}
\setcounter{figure}{0}
\setcounter{table}{0}
\makeatletter
\renewcommand{\theequation}{A\arabic{equation}}
\renewcommand{\thefigure}{A\arabic{figure}}
\setcounter{section}{0}
\renewcommand{\thesection}{\Alph{section}}
\section{Model setup }
\label{sec:model_setup}

We derive here the dynamical equations of the model. We will label locations in the presynaptic layers with Greek indices ($\bm{\alpha, \beta}, \ldots$), and locations in the postsynaptic layer with letters from the end of the Latin alphabet ($\bm{x, y}, \ldots$). 
We call $r(\bm{x}, \tau)$ the firing rate of neurons in the cortical layer, and $r^{\text{ON}}(\bm{\alpha}, \tau)$ and $r^{\text{OFF}}(\bm{\alpha}, \tau)$  the firing rates of ON or OFF neurons in the LGN layers, at time $\tau$.
   
Correlations between the activities of two presynaptic cells  of types $\text{(i)} \in \{  \text{ON, OFF}\}$ and  $\text{(j)} \in \{ \text{ON, OFF}\}$ located at retinotopic positions  $\bm{\alpha}$ and $\bm{\beta}$ are described by a function $C^{(i,j)}(\bm{\alpha}, \bm{\beta}) \equiv C^{(i,j)}(\bm{\alpha} - \bm{\beta})$. Lateral connections between cortical neurons at positions $\bm{x}$ and $\bm{y}$ have synaptic strength given by a function $W(\bm{x}, \bm{y}) \equiv W(\bm{x}- \bm{y})$. We assume both of these to be time-independent over the timescale of orientation map development. 

The quantity of synapses from a presynaptic location $\bm{\alpha}$ to a postsynaptic location $\bm{z}$ is described by an "arbor density" $A(\bm{z}, \bm{\alpha}) \equiv A(\bm{z}-\bm{\alpha})$, which will also be assumed time-independent. The synaptic strength of the $k$-th individual synapse from the afferent at location $\bm{\alpha}$ to the cortical cell at location $\bm{x}$ at time $\tau$ will be called  $s_k^{\text{ON}}(\bm{x},\bm{\alpha},\tau)$ or $s_k^{\text{OFF}}(\bm{x},\bm{\alpha},\tau)$ depending on whether the presynaptic cell is of the ON or OFF type. The index $k$ runs from $1$ to $A(\bm{x}-\bm{\alpha})$. 
  
Finally, let $s^{\text{ON}} (\bm{x}, \bm{\alpha}, \tau) = \sum_{k=1}^{A(\bm{x} - \bm{\alpha})}s_k^{\text{ON}}(\bm{z},\bm{\alpha},\tau)$ and $s^{\text{OFF}} (\bm{x}, \bm{\alpha}, \tau) = \sum_{k=1}^{A(\bm{x} - \bm{\alpha})}s_k^{\text{OFF}}(\bm{z},\bm{\alpha},\tau)$ be the total synaptic strengths at time  $\tau$ from the afferent at location $\bm{\alpha}$ of type ON or OFF to the cortical cell at $\bm{x}$. For ease of notation we will omit the time variable from the arguments of functions.   
    
Applying standard rate dynamics to this model~\cite{Dayan_Abbott01}, we can write that the cortical firing rates evolve according to 
\begin{eqnarray}
\nonumber
\hspace{-10pt} T (\bm{x}) \frac{d r(\bm{x})}{d \tau}  = &-&  r(\bm{x})+ \int \bm{d y} \ W(\bm{x}- \bm{y}) r(\bm{y}) \\  &+& 
+ \sum_{(i) = \text{ON, OFF}} \int  \bm{d \alpha} \  s^{(i)}(\bm{x}, \bm{\alpha}) r^{(i)}(\bm{\alpha})  
\label{firing_rate_one_cell_dynamics}
\end{eqnarray} 
where  $T(\bm{x})$ is a diagonal matrix whose entries are the time scales of neural activity. 

According to Hebb's rule, synapses are strengthened or stabilized if there is temporal correlation between their pre- and postsynaptic patterns of activity. For sufficiently small variations, this can be linearized into the statement that for all $k=1, \ldots, A(\bm{x},\bm{\alpha})$, we have 
\begin{equation}
\label{single_hebbian}
T_{\text{pl}}  \frac{d s_k^{(i)}(\bm{x}, \bm{\alpha})}{d \tau} \propto \ r(\bm{x}) r^{(i)}(\bm{\alpha}),
\end{equation}
where the index $(i)$ distinguishes ON and OFF cells, while $T_{\text{pl}}$ is the time scale for synaptic plasticity. Notice that we are also omitting possible constant terms, included by~\cite{Linsker86} but not essential to the development of selectivity. Assuming symmetry of the two center types, any such constants will disappear when we focus below on the development of the {\it difference} between $s^{\text{ON}}$ and $s^{\text{OFF}}$.

Once Eq.~(\ref{single_hebbian}) is summed over all synapses sharing the same afferent and target cell, we obtain 
\begin{equation}
\label{hebbian}
T_{\text{pl}}  \frac{d s^{(i)}(\bm{x}, \bm{\alpha})}{d \tau} \propto \ A( \bm{x}- \bm{\alpha})  r(\bm{x}) r^{(i)}(\bm{\alpha}).
\end{equation}
The values of $A( \bm{x}-\bm{\alpha})$ are continuous as they represent the local spatial density of arborization from LGN position $\bm{\alpha}$ at cortical position $\bm{x}$.

We will take the time scale $T_{\text{pl}}$ of synaptic plasticity that figures in Eq.~(\ref{single_hebbian}) to be much slower than the time scale of neural activity as given by the entries of $T$ in Eq.~(\ref{firing_rate_one_cell_dynamics}), which seems consistent with experiments~\citep{Kandel}. This allows us to model synaptic development by relying on the steady state of the fast dynamics from Eq.~(\ref{firing_rate_one_cell_dynamics}), which is given by 
\be 
\label{fr_steady_state}
r(\bm{x}) = \sum_{(i) = \text{ON, OFF}} \int \bm{d y} \bm{ d \alpha} \ I(\bm{x}- \bm{y}) s^{(i)}(\bm{y}, \bm{\alpha}) r^{(i)}(\bm{\alpha}),
\ee
where $  I (\bm{x}- \bm{y}) - \int d \bm{z} I(\bm{x}- \bm{z}) W(\bm{z}-\bm{y}) = \delta( \bm{x}- \bm{y})$.
   
Replacing Eq.~(\ref{fr_steady_state}) into Eq.~(\ref{hebbian}) yields
\begin{eqnarray}
\label{2l_hebbian}
&T_{\text{pl}}& \frac{d s^{(i)}(\bm{x}, \bm{\alpha})}{d \tau} \sim A( \bm{x}- \bm{\alpha}) \ \times   \\ && 
\sum_{(j)= \text{ON,OFF}} \int \bm{d y} \bm{d \beta} \  I(\bm{x} - \bm{y}) s^{(j)}(\bm{y,\beta}) r^{(i)}(\bm{\alpha}) r^{(j)}(\bm{\beta}). 
 \nonumber
\end{eqnarray}

We can now average Eq.~(\ref{2l_hebbian}) over a time scale sufficiently longer than the typical time scale of firing-rate dynamics, yet shorter than the typical time scale of synaptic evolution. The averaging leads to the equation
\begin{eqnarray}
\label{unconstrained_dyn}
&&  \frac{d s^{(i)}(\bm{x}, \bm{\alpha})}{d \tau} \sim A( \bm{x}- \bm{\alpha}) \  \times   \\ && 
\sum_{(j)=\text{ON,OFF}} \int \bm{d y} \bm{d \beta} \  I(\bm{x} - \bm{y}) s^{(j)}(\bm{y,\beta})
C^{(ij)}(\bm{\alpha - \beta}), 
 \nonumber
\end{eqnarray}
where $C^{(ij)}_{\bm{\alpha - \beta}} 
= \langle  r^{(i)}_{\bm{\alpha}} r^{(j)}_{\bm{\beta}} 
\rangle $ and we have set the units of time so that $T_{pl} = 1$.
   
Assuming for simplicity symmetry under interchange of ON and OFF, so that $C^{(\text{ON,ON})}=C^{(\text{OFF,OFF})}$, we can transform coordinates to obtain independent equations for  $s^{(S)}\equiv s^{\text{ON}}+s^{\text{OFF}}$ and $s^{(D)}\equiv s^{\text{ON}}-s^{\text{OFF}}$ (superscripts $S$ and $D$ stand for `sum' and `difference', respectively). Our interest is in the development of orientation selectivity via the formation of alternating RF subregions in which ON or OFF LGN inputs, respectively, are dominant. Hence we are interested in development of a pattern in $s^{(D)}$, while $s^{(S)}$ is not expected to form interesting structure. Thus, we will focus on the equation for $s^{(D)}$, which is 
\begin{eqnarray}
    \label{sdequation}
  &&  \frac{d s^{(D)}(\bm{x}, \bm{\alpha})}{d \tau} \sim A( \bm{x}- \bm{\alpha}) \  \times \\ &&  
  \int \bm{d y} \bm{d \beta} \  I(\bm{x} - \bm{y}) s^{(D)}(\bm{y,\beta})
C^{(D)}(\bm{\alpha - \beta}), \nonumber
\end{eqnarray}
where $C^{(D)}=C^{(\text{ON,ON})}-C^{(\text{ON,OFF})}$ is the difference between same-center-type and opposite-center-type correlations.

It is characteristic of Hebbian rules that synaptic strengths tend to increase without limit~\cite{Dayan_Abbott01}. Ref.~\cite{miller94model} modeled the biological mechanisms for saturation by including an upper bound and a zero lower bound for all synaptic strengths ($0\leq s^{(i)} \leq s_{\max}$), which becomes a limit on $s^{(D)}$ of $- s_{\max} \leq s^{(D)} \leq s_{\max}$. This turns Eq.~(\ref{sdequation}) into a nonlinear equation. However, we imagine development starting from an initial condition in which there are roughly equal strengths of ON and OFF innervation throughout the receptive field, and thus in which the values of $s^{(D)}$ are small random perturbations about 0. We can always assume the synaptic weight bounds large enough so that the principal features of the $s^{(D)}$ dynamics are established before the bounds are saturated (as in~\cite{Linsker86, miller94model}). Once the bounds are reached, they will simply capture and preserve the existing weight structure with little subsequent change. 
   
We can therefore extract the long-term behavior of the synaptic weights simply by analyzing the properties of the time-evolution operator  in the linear regime.  As per Eq.~(\ref{sdequation}), this will be, in a first approximation, the integral operator characterized by the kernel 
\begin{equation}
\label{kernelequation}
K^{(D)}(\bm{x}, \bm{\alpha}; \bm{y}, \bm{\beta}) = A(\bm{x}- \bm{\alpha}) I(\bm{x} - \bm{y}) C^{(D)}(\bm{\alpha - \beta}). 
\end{equation}
 
We now incorporate the fact that correlation-based development is competitive. In this framework, indeed, a mechanism of the order of competition is necessary for neurons to become selective for certain features. Without competition, all synapses onto a cell could grow to their maximum possible value, eliminating all selectivity save the retinotopic selectivity embodied in the arbor density. 
As discussed in the main text, we model competition by conserving the total strength of synapses projected by each arbor:
\begin{eqnarray}
 \label{conservation}
 \frac{d}{d \tau}  \int {\bm{d x}}\  s^{\text{ON}}_{\bm{x, \alpha} } =
 \frac{d}{d \tau}  \int {\bm{d x} }\ s^{\text{OFF}}_{\bm{x, \alpha} } = 0  \ \ \ \forall {\bm{\alpha}},
\end{eqnarray}
where the arguments of functions have been written as subscripts for the sake of compactness. 

Together, these imply
\begin{equation}
    \label{sdconservation}
     \frac{d}{d \tau}  \int {\bm{d x}}\  s^{\text{D}}_{\bm{x, \alpha} } =
 \frac{d}{d \tau}  \int {\bm{d x} }\ s^{\text{S}}_{\bm{x, \alpha} } = 0  \ \ \ \forall {\bm{\alpha}}.
\end{equation}

Henceforth we focus only on the development of $s^{(D)}$. We will drop the `D` superscript, simply writing $s$ for $s^{(D)}$, $C$ for $C^{(D)}$, and $K$ for $K^{(D)}$ (Eq.~\ref{kernelequation}). Including the constraint, Eq.~\ref{sdconservation}, the equation we study is 
\begin{eqnarray}
\label{dynamics}
 \frac{d s(\bm{x}, \bm{\alpha})}{d \tau} &=&  \int \bm{d y} \bm{d \beta} \ K(\bm{x}, \bm{\alpha}; \bm{y}, \bm{\beta})  s(\bm{y,\beta})  
 \\  && \hspace{-30pt} 
 -\frac{A(\bm{x}- \bm{\alpha})}{\int \bm{d z} A(\bm{z}- \bm{\alpha})}
   \int \bm{d y} \bm{d q}  \bm{d \beta} \ K(\bm{q}, \bm{\alpha}; \bm{y}, \bm{\beta})  s(\bm{y,\beta}).
\nonumber
\end{eqnarray}
  
To completely specify the model, it only remains to choose a form for the functions $A$, $I$ and $C$. The assumption of  monotonically decreasing functions reflects a principle of modeling economy for $I$, and is suggested by the general decay of arborizations with distance for $A$ and by the experimental results discussed in the Introduction for $C$.We  take  them for definiteness to have  Gaussian dependencies on distances: 
\begin{eqnarray} 
\label{gaussian_choice_arbor}
A(\bm{x}-\bm{\alpha}) &\propto& e^{ - \frac{(\bm{x}-\bm{\alpha})^2}{2 \rho^2} },
 \hspace{30pt}
\\
\label{gaussian_choice_corr}
C(\bm{\alpha}-\bm{ \beta}) &\propto& e^{ - \frac{(\bm{\alpha}-\bm{\beta})^2}{2 \zeta^2} }, 
\hspace{30pt} 
\\
\label{gaussian_choice_interaction}
I (\bm{x}- \bm{y}) &\propto& e^{ - \frac{(\bm{x}-\bm{y})^2}{2 \eta^2} },
\end{eqnarray}
where the arbor radius $\rho$, the interaction length $\eta$ and the correlation length $\zeta$ are the three characteristic length-scales of the model. 

\section{Hermitian formulation of the constrained problem}
\label{sec:homeostatic} 

The homeostatic constraint mechanism modeled by Eq.~(\ref{conservation}) conserves the total projection strength from each presynaptic cell. We start by illustrating how this constraint can be incorporated into the theory; namely, by adding to the Hebbian law Eq.~(\ref{single_hebbian}) a suitable leak term  $\epsilon^{\text{ON}}_{\bm{\alpha}}$. This yields
\begin{equation}
\label{single_hebbian_with_leak}
\frac{d s_k^{(i)}(\bm{x}, \bm{\alpha})}{d \tau} \propto \ - \epsilon^{(i)}_{\bm{\alpha}} + r(\bm{x}) r^{(i)}(\bm{\alpha}) 
\end{equation}
Here, $\epsilon^{\text{ON}}_{\bm{\alpha}}$ and $\epsilon^{\text{OFF}}_{\bm{\alpha}}$  are unspecified quantities that  will be defined in such a way as to implement the conservation constraints Eq.~(\ref{conservation}) of the model. 
 
Summing Eq.~(\ref{single_hebbian_with_leak}) over all synapses with the same afferent  and performing the averaging over time as done for Eq.~(\ref{2l_hebbian}), we obtain 
\begin{align}
\label{adding_constraint_1}
\frac{d}{d \tau} s^{\text{ON}}_{\bm{x, \alpha}} &= - \epsilon^{\text{ON}}_{\bm{\alpha}} A_{\bm{x-\alpha}} 
+ \\ \nonumber 
+ A_{\bm{x-\alpha}}  &\sum_{\bm{y, \beta}} I_{\bm{x-y}} \left[C^{	\text{ON, ON}}_{\bm{\alpha - \beta}} s^{\text{ON}}_{\bm{y, \beta}} + 
C^{\text{ON, OFF}}_{\bm{\alpha - \beta}} s^{\text{OFF}}_{\bm{y, \beta}} \right], 
\end{align}

\begin{align}
\label{adding_constraint_2}
\frac{d}{d \tau} s^{\text{OFF}} _{\bm{x, \alpha}} &= - \epsilon^{\text{OFF}}_{\bm{\alpha}}A_{\bm{x-\alpha}}
+ \\  \nonumber 
+ A_{\bm{x-\alpha}} &\sum_{\bm{y, \beta}} I_{\bm{x-y}} 
\left[C^{\text{OFF, OFF}}_{\bm{\alpha - \beta}} s^{\text{OFF}}_{\bm{y, \beta}} + 
C^{\text{OFF, ON}}_{\bm{\alpha - \beta}} s^{\text{ON}}_{\bm{y, \beta}} \right], 
\end{align}
 
The crucial step for all that follows is to render the operation in Eqs.~(\ref{adding_constraint_1}-\ref{adding_constraint_2}) symmetric, which re-defines the time-evolution in terms of Hermitian operators. We do so by defining 
\be
t^{\text{ON}}_{\bm{x,\alpha}} = \frac{s^{\text{ON}}_{\bm{x, \alpha}}}{ \sqrt{A_{\bm{x-\alpha}} }} 
\hspace{30pt} 
t^{\text{OFF}}_{\bm{x,\alpha}} = \frac{s^{\text{OFF}} _{\bm{x, \alpha}}  }{ \sqrt{A_{\bm{x-\alpha}} }}
\ee
so that Eqs.~(\ref{adding_constraint_1}-\ref{adding_constraint_2}) become
\begin{widetext}
\bea 
\label{expl_constr_1}
 \frac{d}{d\tau} t^{\text{ON}} _{\bm{x, \alpha}} = \sqrt{A_{\bm{x-\alpha}} } \left\{ 
 - \epsilon^{\text{ON}}_{\bm{\alpha}} + 
 \sum_{\bm{y, \beta}} I_{\bm{x-y}} \left[C^{\text{ON, ON}}_{\bm{\alpha - \beta}} \sqrt{A_{\bm{y - \beta}} }  \ t^{\text{ON}}_{\bm{y, \beta}} + 
C^{\text{ON, OFF}}_{\bm{\alpha - \beta}} \sqrt{A_{\bm{y - \beta}} }  \ t^{\text{OFF}}_{\bm{y, \beta}} \right]   \right\} , 
\\ \frac{d}{d\tau} t^{\text{OFF}} _{\bm{x, \alpha}} = \sqrt{A_{\bm{x-\alpha}} } \left\{ - \epsilon^{\text{OFF}}_{\bm{\alpha}} + \sum_{\bm{y, \beta}} I_{\bm{x-y}} \left[C^{\text{OFF, OFF}}_{\bm{\alpha - \beta}} \sqrt{A_{\bm{y - \beta}} }  \ t^{\text{OFF}}_{\bm{y, \beta}} + C^{\text{OFF, ON}}_{\bm{\alpha - \beta}} \sqrt{A_{\bm{y - \beta}} }  \ t^{\text{ON}}_{\bm{y, \beta}} \right]  \right\}. 
\label{expl_constr_2}
\eea
\end{widetext} 

It will be convenient to regard the functions $t^{\text{ON}}$  and $t^{\text{OFF}}$  as vectors in a Hilbert space, so that (in bra/ket notation) $ t^{\text{ON}} _{\bm{x, \alpha}}  = \langle \bm{x, \alpha} | t^{\text{ON}} \rangle $ and $ t^{\text{OFF}} _{\bm{x, \alpha}}  = \langle \bm{x, \alpha} | t^{\text{OFF}} \rangle $. 
  
Since ON and OFF cells are subjected to the same inputs from the retina, we may assume $C^{\text{ON,ON}} = C^{\text{OFF,OFF}}$; using this and the fact that $C^{\text{ON,OFF}} = C^{\text{OFF,ON}}$, we rewrite Eqs.~(\ref{expl_constr_1}-\ref{expl_constr_2}) in terms of the sole operators $\hat{L}^s$ and $\hat{L}^d$, defined as follows (hats on the names of operators distinguish them from ordinary variables): 
\begin{eqnarray}
\hspace{-30pt}
\langle \bm{x, \alpha} |  \hat{L}^s   | \bm{y, \beta}  \rangle =  \sqrt{A_{\bm{x-\alpha}} } I_{\bm{x-y}}  C^{\text{ON, ON}}_{\bm{\alpha - \beta}} \sqrt{A_{\bm{y - \beta}} }, 
\\
\hspace{-30pt}
\langle \bm{x, \alpha} |  \hat{L}^d   | \bm{y, \beta}  \rangle =  \sqrt{A_{\bm{x-\alpha}} } I_{\bm{x-y}}  C^{\text{ON, OFF}}_{\bm{\alpha - \beta}} \sqrt{A_{\bm{y - \beta}} }, 
\end{eqnarray}
yielding
\begin{eqnarray}
\label{tderiv1}
\hspace{-25pt}
 \frac{d}{d\tau}  | t^{\text{ON}} \rangle  =   \hat{L}^s  | t^{\text{ON}} \rangle   
+\hat{L}^d | t^{\text{OFF}} \rangle   -\hat{\mathcal{E}}^{\text{ON}} | a \rangle,  \\ 
\label{tderiv2}
\hspace{-25pt}
\frac{d}{d\tau}  | t^{\text{OFF}} \rangle  =   \hat{L}^s  | t^{\text{OFF}} \rangle   
+\hat{L}^d | t^{\text{ON}} \rangle   - \hat{\mathcal{E}}^{\text{OFF}} | a \rangle.   
\end{eqnarray}
 
Here the vector $| a  \rangle $ is defined by  $ \langle \bm{x, \alpha} | a \rangle = \sqrt{A_{\bm{x-\alpha}}} $, while the operators $\hat{\mathcal{E}}^{\text{ON}}$ and $\hat{\mathcal{E}}^{\text{OFF}}$ 
have the form 
\be 
\label{epsilons_c}
\hat{\mathcal{E}}^{\text{ON}}  = \sum_{\bm{\beta}} \epsilon^{\text{ON}}_\beta \hat{P}_{\bm{\beta}} \hspace{35pt} \hat{\mathcal{E}}^{\text{OFF}}  = \sum_{\bm{\beta}} \epsilon^{\text{OFF}}_\beta \hat{P}_{\bm{\beta}}, 
\ee 
where $\hat{P}_{\bm{\beta}}$ is the operator that effects projection into the subspace with basis $\{ | \bm{y, \beta} \rangle \}_{\bm{y}}$, that is, 
\be
\label{proj_op_decomposition}
 \hat{P}_{\bm{\beta}} \equiv \sum_{\bm{y}} | \bm{y, \beta}  \rangle \langle \bm{y, \beta} |.
 \ee
 
The expressions for $\epsilon^{\text{ON}}$, $\epsilon^{\text{OFF}}$ can be found from the conservation laws Eq.~(\ref{conservation}), 
which may be rewritten in the form 
\be
 \frac{d}{d \tau}  \sum_{\bm{y} } \  \sqrt{A_{\bm{y-\beta}}} \ \ t^{ON,OFF}_{\bm{y,\beta}} = 0  \ \ \ \forall {\bm{\beta}}
\ee
or rather
\begin{eqnarray}
\label{2l_constraint1}
 \frac{d}{d\tau}  \langle a | \hat{P}_{\bm{\beta}} | t^{\text{ON}} \rangle &=& 0 \ \  \ \forall \ {\bm{\beta} },
\\
\label{2l_constraint2}
 \frac{d}{d\tau}  \langle a | \hat{P}_{\bm{\beta}} | t^{\text{OFF}} \rangle &=& 0 \ \  \ \forall \ {\bm{\beta} }.
\end{eqnarray}

Substituting Eqs.~(\ref{tderiv1}-\ref{tderiv2})
into the two constraints (\ref{2l_constraint1},\ref{2l_constraint1}), and using the expressions (\ref{epsilons_c}) for $\mathcal{E} ^{\text{ON}} $, $\mathcal{E} ^{\text{OFF}}$, we finally arrive at 
\begin{eqnarray}
\epsilon^{\text{ON}}_{\bm{\beta} } &=&  \frac{ \langle a | \hat{P}_{\bm{\beta}} L^s | t^{\text{ON}} \rangle  + 
\langle a | \hat{P}_{\bm{\beta}} L^d | t^{\text{OFF}} \rangle   }{\langle a | \hat{P}_{\bm{\beta}}  | a \rangle},
\\
\epsilon^{\text{OFF}}_{\bm{\beta} } &=&  \frac{ \langle a | \hat{P}_{\bm{\beta}} L^s | t^{\text{OFF}} \rangle  + 
\langle a | \hat{P}_{\bm{\beta}} L^d | t^{\text{ON}} \rangle   }{ \langle a | \hat{P}_{\bm{\beta}}  | a \rangle}.
\end{eqnarray}
   
For the difference $s_{\bm{x,\alpha}} \equiv s^{\text{ON}}_{\bm{x, \alpha} } -  s^{\text{OFF}}_{\bm{x, \alpha} }$ we have  
\be
\label{t_to_s}
s_{\bm{x,\alpha}} = \sqrt{A_{\bm{x-\alpha}} } \left(
t^{\text{ON}}_{\bm{x, \alpha} } -  t^{\text{OFF}}_{\bm{x, \alpha} } \right),
\ee
so that
we must proceed to compute the time evolution of $| t \rangle \equiv | t^{\text{ON}} - t^{\text{OFF}} \rangle $. From Eqs.~(\ref{tderiv1}-\ref{tderiv2}), we find  
\begin{equation}
\label{time_ev}
 \frac{d}{d\tau}  | t \rangle = \left[ \bm{1} - \sum_{\bm{\beta}} \frac{\hat{P}_{\bm{\beta}}  |a\rangle \langle a| \hat{P}_{\bm{\beta}}  }{ \langle a| \hat{P}_{\bm{\beta}}  | a \rangle } \right] \hat{L} |t \rangle, 
\end{equation}
where $\bm{1}$ stands for the identity operator and we have defined $\hat{L} \equiv \hat{L}^s - \hat{L}^d$. 

\vspace{20pt}

\section{Projection operators}
\label{sec:projection_operators}

To rewrite Eq.~(\ref{time_ev}) in a more transparent form, we define the single-arbor ket $|a_{\bm{\beta}}   \rangle = \hat{P}_{\bm{\beta}}  |a\rangle $, with elements $\langle \bm{x, \alpha } | a_{\bm{\beta}}\rangle  = \delta_{\bm{\alpha , \beta}} 
\sqrt{A_{\bm{x - \alpha}}}$. Notice that the operator $\hat{P}_{\bm{\beta}}$ of Eq.~(\ref{proj_op_decomposition})
is orthogonal and therefore, being also a projection, it is self-adjoint. Using this fact, as well as the idempotence of $\hat{P}_{\bm{\beta}}$, we obtain
\be \label{constrained}
\frac{d}{d\tau}  | t \rangle = \left[ \bm{1} - \sum_{\bm{\beta}} \frac{|a_{\bm{\beta}}\rangle \langle a_{\bm{\beta}}| }{ \langle a_{\bm{\beta}}| | a_{\bm{\beta}} \rangle } \right] \hat{L} |t \rangle.  
\ee
 
Defining $\hat{P} = \bm{1} - \sum_{\bm{\beta}} \frac{|a_{\bm{\beta}}\rangle \langle a_{\bm{\beta}}| }{ \langle a_{\bm{\beta}}| a_{\bm{\beta}} \rangle }$, 
we have from Eq.~(\ref{constrained})
\be 
\label{decompose}
\frac{d}{d\tau}  | t \rangle = \hat{P} \hat{L} | t \rangle = \hat{P}\hat{L}\hat{P} | t \rangle  + \hat{P}\hat{L} (\bm{1}- \hat{P}) | t \rangle.  
\ee
 
Since we are interested in calculating the final outcome of development, we must focus on the long-term behavior of this dynamics. To do so, we notice that the components of $|t \rangle$  projected away by $\hat{P}$ cannot be made to grow by the constrained Hebbian dynamics of Eq.~(\ref{decompose}). Therefore positive eigenvalues leading to exponential growth must be found in the space in which $\hat{P}$ is projecting, and after waiting a sufficiently long time, one may always approximate the state of the system as contained in that space. 

Noting this, we can drop the last term in Eq.~(\ref{decompose}) and write simply 
\be
\label{main}
\frac{d}{d\tau}  | t \rangle = \hat{P}\hat{L}\hat{P} | t \rangle \equiv  \hat{L}^p \ | t \rangle
\ee

Finally notice that, since the arbor density $A(\bm{r})$ has the meaning of a density, we can define it as being properly normalized so that $\int d\bm{r} A(\bm{r})=1 $, i.e. $\langle a_{\bm{\beta}}| | a_{\bm{\beta}} \rangle =1$. This allows to remove the denominators in the definition of the projection operator $\hat{P}$, which becomes simply
\be 
\hat{P} = \bm{1} - \sum_{\bm{\beta}} | a_{\bm{\beta}}\rangle \langle a_{\bm{\beta}}|
\label{simplified_projection_operator}
\ee

The principal eigenspace of the operator $L^p$ of Eq.~\ref{main} defined in terms of Eq.~\ref{simplified_projection_operator} is thus the object of our interest as it will determine the fastest-growing modes of the system. 

\bibliography{library}

\newpage
  
\onecolumngrid
\setcounter{page}{0}
\pagenumbering{arabic}
\setcounter{page}{1}
\clearpage
 
\begin{center}
\textbf{SUPPLEMENTAL MATERIAL}
\end{center}

\setcounter{equation}{0}
\setcounter{figure}{0}
\setcounter{table}{0}
\setcounter{page}{1}
\setcounter{section}{0}
\makeatletter
\renewcommand{\thesection}{S-\Roman{section}}
\renewcommand{\theequation}{S\arabic{equation}}
\renewcommand{\thefigure}{S\arabic{figure}}

\medskip 

\section{Properties of the Hermitianized operator} 
\label{sec:properties}
 
We now analyze the basic properties of the time-evolution operator 
$\hat{L}^p$ of Eq.~(\ref{main}), including the form of its matrix elements (A) in real space and (B) after Fourier transforming 
in the cortical variables, (C) its positive semidefiniteness, (D) the general structure of its spectrum, (E) its commutation properties with translation and rotation operators, (F) its symmetry with respect to parity, complex conjugation, and their combination; (G) we finally write down the exact diagonalization of the unconstrained operator $L$, which will serve as the starting point for studying the properties of $\hat{L}^p$ in greater detail.
    
\subsection{Matrix elements}

The matrix elements of the unconstrained operator $\hat{L}$ appearing in Eq.~(\ref{time_ev}) are, in the $|\bm{ x, \alpha} \rangle $  basis,
\begin{equation}
\label{L_mat_element}
L(\bm{x,\alpha};\bm{y,\beta}) \equiv \langle \bm{x, \alpha} |  L   | \bm{y, \beta}  \rangle =  \sqrt{A_{\bm{x-\alpha}} } I_{\bm{x-y}}  C_{\bm{\alpha - \beta}} \sqrt{A_{\bm{y - \beta}} },  
\end{equation}
where $C_{\bm{\alpha - \beta}} = C^{\text{ON, ON}}_{\bm{\alpha - \beta}}  - C^{\text{ON, OFF}}_{\bm{\alpha - \beta}} $. 

The generic matrix element of the operator $\hat{L}^p $  of Eq.~(\ref{main}) is written, using 
\ref{simplified_projection_operator},  as
\begin{eqnarray}
\nonumber  
&& L^p(\bm{x},\bm{\alpha}; \bm{y},\bm{\beta}) 
= \langle \bm{x, \alpha} |  \hat{L}^p   | \bm{y, \beta}  \rangle =  
\int \bm{d x_1 d \alpha_1  d x_2 d \alpha_2} \bigg[ \delta(\bm{x- x_1}) \delta(\bm{\alpha- \alpha_1}) 
\\ 
\nonumber 
&-& 
\sqrt{A(\bm{x- \alpha}) }\delta( \bm{\alpha - \alpha_1}) \sqrt{A(\bm{x_1 - \alpha_1} )} \bigg]   L(\bm{x_1, \alpha_1; x_2, \alpha_2}) \times \\ && \times
\bigg[ \delta(\bm{x_2- y}) \delta(\bm{\alpha_2- \beta}) - \sqrt{A(\bm{x_2 - \alpha_2 }) } \delta( \bm{\alpha_2 - \beta}) \sqrt{A(\bm{y  - \beta})} \bigg]. \hspace{30pt}
\label{Lp_first_real_space_matrix_elements}
\end{eqnarray}
 
The natural variables in which to express a RF are the relative coordinates $\bm{r} = \bm{\alpha - x}$, and we will abuse the notation by writing $L^p(\bm{x},\bm{r}; \bm{y},\bm{s})  \equiv L^p(\bm{x},\bm{\alpha - x}; \bm{y},\bm{\beta - y})$ and $L(\bm{x},\bm{r}; \bm{y},\bm{s})  \equiv L(\bm{x},\bm{\alpha - x}; \bm{y},\bm{\beta - y})$. Integrating out the delta functions in Eq.~(\ref{Lp_first_real_space_matrix_elements}) we obtain
\begin{eqnarray}
\label{Lp_matrix_element}
L^p(\bm{x},\bm{r}; \bm{y},\bm{s}) = L(\bm{x},\bm{r}; \bm{y},\bm{s})  + S(\bm{x},\bm{r}; \bm{y},\bm{s})  + T(\bm{x},\bm{r}; \bm{y},\bm{s})  + \tilde{T} (\bm{x},\bm{r}; \bm{y},\bm{s}),  \end{eqnarray}
On the RHS of this equation, the first term is obtain with a change of variables in the arguments of Eq.~(\ref{L_mat_element}), leading to 

\be 
\label{unconstrained_matr_el}
L (\bm{x},\bm{r}; \bm{y}, \bm{s}) = \sqrt{A(r) } \sqrt{A(s ) } I( \bm{x- y}) C(\bm{x- y + r - s}).
\ee

while the last three terms are given by

\begin{eqnarray}
\label{O1}
\hspace{-.7cm} \frac{S (\bm{x},\bm{r}; \bm{y},\bm{s}) }{ \sqrt{A(r)A(s)} } &=&
\int \bm{d s}_1  \bm{d s}_2 \sqrt{A(s_1) A(s_2)}  
L(\bm{x + r - s_1 , s_1; y + s - s_2, s_2})
\\ \label{O2}
\hspace{-.7cm}  T(\bm{x},\bm{r}; \bm{y},\bm{s})  &=& - \sqrt{A(r)} \int \bm{d u} \sqrt{A(u)}  L(\bm{x + r - u, u; y, s}) \\ 
\label{Otilde2}
 \hspace{-.7cm}  \tilde{T} (\bm{x},\bm{r}; \bm{y},\bm{s})  &=& 
 - \sqrt{A(s)} \int \bm{d u} \sqrt{A(u)}  L(\bm{x, r; y + s - u, u}).
 \end{eqnarray}

In the special case where the functions $I(\bm{x}-\bm{y})$ and $C(\bm{\alpha}-\bm{\beta})$ are even under parity, it is seen from Eq.~(\ref{unconstrained_matr_el}) that  $\hat{L}$ becomes symmetric under swapping of LGN and cortical coordinates, and it follows that $\hat{S}=\hat{S}^T$, $\hat{\tilde{T}}=\hat{T}^T$. 
\subsection{Cortical Fourier Transform}
 
Noting the translation invariance of $\hat{L}$ in Eq.~(\ref{unconstrained_matr_el}) with respect to the cortical location variable, we can define 
\be
 L(\bm{x,r; y,s}) \equiv 
\int \frac{\bm{d \omega}}{(2 \pi)^2} e^{- i \bm{\omega (x - y)}} L(\bm{r,s; \omega}). 
\ee
and similarly for $\hat{L}^p$, from the translation invariance seen in Eq.~(\ref{Lp_first_real_space_matrix_elements}). This means that the eigenfunctions of $\hat{L}^p$ in Eq.~(\ref{Lp_matrix_element}) will be of the form $e^{i \bm{ \omega x}} \psi_\omega (\bm{r})$, where  
$e^{i \bm{ \omega x}} $ describes an oscillation across the cortical coordinate $\bm{x}$ and $\psi_\omega(\bm{r})$  describes the RF as a function of the LGN position $\bm{\alpha}$ relative to $\bm{x}$, i.e. $\bm{r}=\bm{ \alpha - x }$.  

The function $\psi_\omega(\bm{r})$ is complex, and we will write it in the form of real functions as $\psi_\omega(\bm{r}) = 
\psi^R_\omega(\bm{r}) + i \psi^L_\omega(\bm{r})$. Then this eigenfunction corresponds to the real functions $
\cos(\bm{\omega x} +\phi) \psi^R_\omega(\bm{r}) + \sin(\bm{\omega x } +\phi) \psi^R_\omega(\bm{r}) $ 
for arbitrary phase $\phi$. 

We will refer to the spatial frequency vector $\bm{\omega}$ as the cortical wavevector and to its modulus as the cortical wavenumber. The Fourier transform of the constrained operator, Eq.~(\ref{Lp_first_real_space_matrix_elements}), which determines the RF eigenfunctions $\psi_\omega(\bm{r})$, is then
\begin{eqnarray}
L^p (\bm{r,s; \omega}) =  L(\bm{r, s ; \omega}) - 
\sqrt{A(r)} \int \bm{d r_1} \sqrt{A(r_1)}  L(\bm{r_1, s; \omega}) e^{ - i \bm{\omega( r - r_1)} } \nonumber\\
- 
\sqrt{A(s)} \int \bm{d s_1} \sqrt{A(s_1)}  L(\bm{r, s_1; \omega}) e^{ - i \bm{\omega( s_1 -s )} }  \nonumber
\\ + \sqrt{A(r) A(s)} \int dr_1 ds_1 \sqrt{A(r_1) A(s_1)}  L(\bm{r_1, s_1; \omega})e^{ - i \bm{\omega ( r - s - r_1 + s_1)}}.
\label{LP_gen_element}
\end{eqnarray}
  
With the choice~(\ref{gaussian_choice_corr},\ref{gaussian_choice_interaction}) for the interaction and correlation functions, the transform of Eq.~(\ref{unconstrained_matr_el}) reads 
\be 
L(\bm{r,s;\omega}) \sim L^{u}(\bm{r,s; \mu}) \exp\left[ - \frac{\omega^2}{2 \Omega^2} - i \frac{\eta^2}{\mu^2} \bm{\omega (r -s )}\right], \label{fourier-transformed-unconstrained}
\ee
where we have neglected an overall prefactor that can be absorbed in the definition of time. The "effective length" $\mu$ and "cutoff wavenumber" $\Omega$ in Eq.~(\ref{fourier-transformed-unconstrained}) are given by
 \be 
\label{mu-and-omega}
 \mu^2=\eta^2 + \zeta^2 \hspace{70pt}  \Omega^2 = \frac{1}{\eta^2} + \frac{1}{\zeta^2},
 \ee
and $L^{u}(r,s; \mu) = \sqrt{A(r) A(s) } e^{- \frac{(\bm{r-s})^2}{2 \mu^2}}$, with the apex $u$ standing for "unconstrained". We may refer to the parameter $\mu$ as to the correlation-interaction length, as it is a Pythagorean combination of the two "intra-layer" length scales of the problem.
 
We now (a) substitute the Fourier-transformed matrix element of $\hat{L}$ as per  Eq.~(\ref{fourier-transformed-unconstrained})
into the expression~(\ref{LP_gen_element}) for the matrix element of $\hat{L}^p$, (b) insert a specific (Gaussian) assumption for the arbor density function $A(r) = \frac{1}{2 \pi \rho^2 } \exp\left( - \frac{r^2}{2 \rho^2}\right)$, and (c) perform the integration over all intermediate space variables. 
 
We thus arrive at decomposing the constrained operator of Eq.~(\ref{Lp_first_real_space_matrix_elements}) into
\begin{equation} \hat{L}^p = 
\hat{L} + \hat{S} + \hat{T} + \hat{T}^{\dagger}, 
\label{form1}
\end{equation} 
where in the coordinate representation these are 
\begin{eqnarray} 
\label{form2}
\hspace{-20pt} 
L 
(\bm{r,s}; \bm{\omega}) &=& \exp\left[
-\frac{ \omega^2}{2 \Omega^2} - i \frac{\eta^2}{\mu^2} \bm{\omega} (\bm{r} - \bm{s}) - \frac{r^2 + s^2}{4 \rho^2} 
- \frac{(\bm{r} - \bm{s})^2}{2 \mu^2} \right];
\\ 
\label{form3}
\hspace{-20pt} 
S (\bm{r,s}; \bm{\omega}) &=& 
\frac{\mu^2}{ \mu^2 + 2 \rho^2 } \exp\left[ 
-\frac{ \omega^2}{2 \Omega^2} - \frac{\rho^2 \zeta^4 \omega^2}{\mu^2(\mu^2 + 2 \rho^2)} - i \bm{\omega} (\bm{r} - \bm{s}) - \frac{r^2 + s^2}{4 \rho^2}  \right];
\\
 \label{form4}
\hspace{-20pt} 
\nonumber
T (\bm{r,s}; \bm{\omega}) &=&  -
 \frac{\mu^2}{\mu^2 + \rho^2 } \exp\Big[ -\frac{ \omega^2}{2 \Omega^2} - \frac{\rho^2 \zeta^4 \omega^2 }{2 \mu^2 ( \mu^2 + \rho^2)} - \frac{1}{2} \left( 
\frac{1}{2 \rho^2 }   +  \frac{1}{\rho^2 + \mu^2} \right) r^2 \hspace{10pt} 
\\ 
&-& \frac{s^2}{4 \rho^2}  
 - i \bm{\omega} \left( \frac{\rho^2 + \eta^2}{\rho^2 + \mu^2} \bm{r}  - \bm{s}\right) \Big];
\\
 \label{form5}
\hspace{-20pt} 
\nonumber
T^\dagger(\bm{r,s};\bm{\omega}) &=& T^*(\bm{s,r};\bm{\omega}) 
\end{eqnarray}
and the additive constraint operators $\hat{S}$ and $\hat{T}$ have fully separable matrix elements. 
 
\subsection{Positive semidefiniteness} 
\label{two_layers_positive_semid}

It will be useful to rely on the positive semidefiniteness of the Fourier-projected operator $\hat{L}^p(\omega)$ for any given wavenumber $\omega$. 

Consider Eq.~(\ref{LP_gen_element}), and suppose we regard the direction of the wavevector $\bm{\omega}$ as fixed (in the following, with no loss of generality, we will take it to be parallel to the x-axis). We can then write 
\be 
\label{proj_form}
\hat{L}^p (\omega) = \Big( 1 - | a_{\omega} \rangle \langle a_\omega | \Big) \hat{L}(\omega) \Big( 1 - | a_\omega \rangle \langle a_\omega | \Big),
\ee
where $\langle \bm{r} | a_\omega \rangle = \sqrt{A(r)} e^{- i \omega r_x}$. 
 
Thus, even for a given wavenumber $\omega$, the constrained operator is nothing but the unconstrained operator 
sandwiched between two identical projection operators.

The unconstrained operator $\hat{L}$ is clearly positive definite. This follows from the fact that it is Hermitian with an all-positive kernel; we will see explicitly that its eigenvalues are all positive but dense in a neighborhood of zero. Recall now the following lemma: if $\hat{O}$ is a positive-definite operator in a linear space and $\hat{P}$ a projection operator on some subspace, then $\hat{P}\hat{O} \hat{P}$ is positive-semidefinite -- from which it follows that $\hat{L}^{p}$ is positive-semidefinite.

\subsection{Long-term dynamics } 
\label{sec:long-term-dyn}

Once the operator is diagonalized for an arbitrary cortical wavenumber, we expect to find the eigenvalues from a series of possibly overlapping bands, where each given band corresponds to a set of eigenfunctions with varying $\omega$. Different bands 
may come from different sets of eigenstates characterized by discrete numbers that we may term quantum numbers. 
As we will show, for low wavenumbers these correspond to different rotational eigenstates. We will call "principal band" of the spectrum the one that contains the principal mode, which drives the long-term dynamics. 

Call $\Lambda_M(\omega)$ the wavenumber-dependent eigenvalues of the principal band, and $\omega_M$ the position of its (possibly broad) maximum as a function of $\omega$, corresponding to the fastest-growing eigenspace. In the following, we will be interested in the principal eigenspace of the operator $L^p $, as this determines the fastest growing modes. We can gain insight on long-term behavior by focusing on the principal band, and on the RFs it represents. 
  
Indeed, the developmental process for a given band has a similar effect as filtering with a spatially isotropic bandpass with respect to cortical wavenumber. The function $\Lambda_M(\omega)$ can be interpreted as the corresponding filter profile, and the location of the maximum of this filter may depend in nontrivial ways on the model's parameters. 
 
If the system lies in parameter space at a point such that $\omega_M=0$ , the dynamics will tend to flatten out any spatial inhomogeneity in the initial condition. If $\omega_M > 0$, on the other hand, the long-term RF will vary spatially on a scale $\sim 1/\omega_M$.  Since, as will be seen, we have a broad maximum of nearly optimal wavevector, we may expect local but no long-range periodicity. Anisotropies in the initial conditions can also be magnified by the dynamics. 

The evolution of the RF at any given point in the cortex, finally, may cancel or emphasize whatever degree of orientation selectivity is possessed by the initial condition, depending on the structure of the eigenspace associated to the principal mode.

\subsection{Symmetries of the system: translations and rotations}
    
\begin{figure}[htp]
\includegraphics[width=0.6\textwidth]{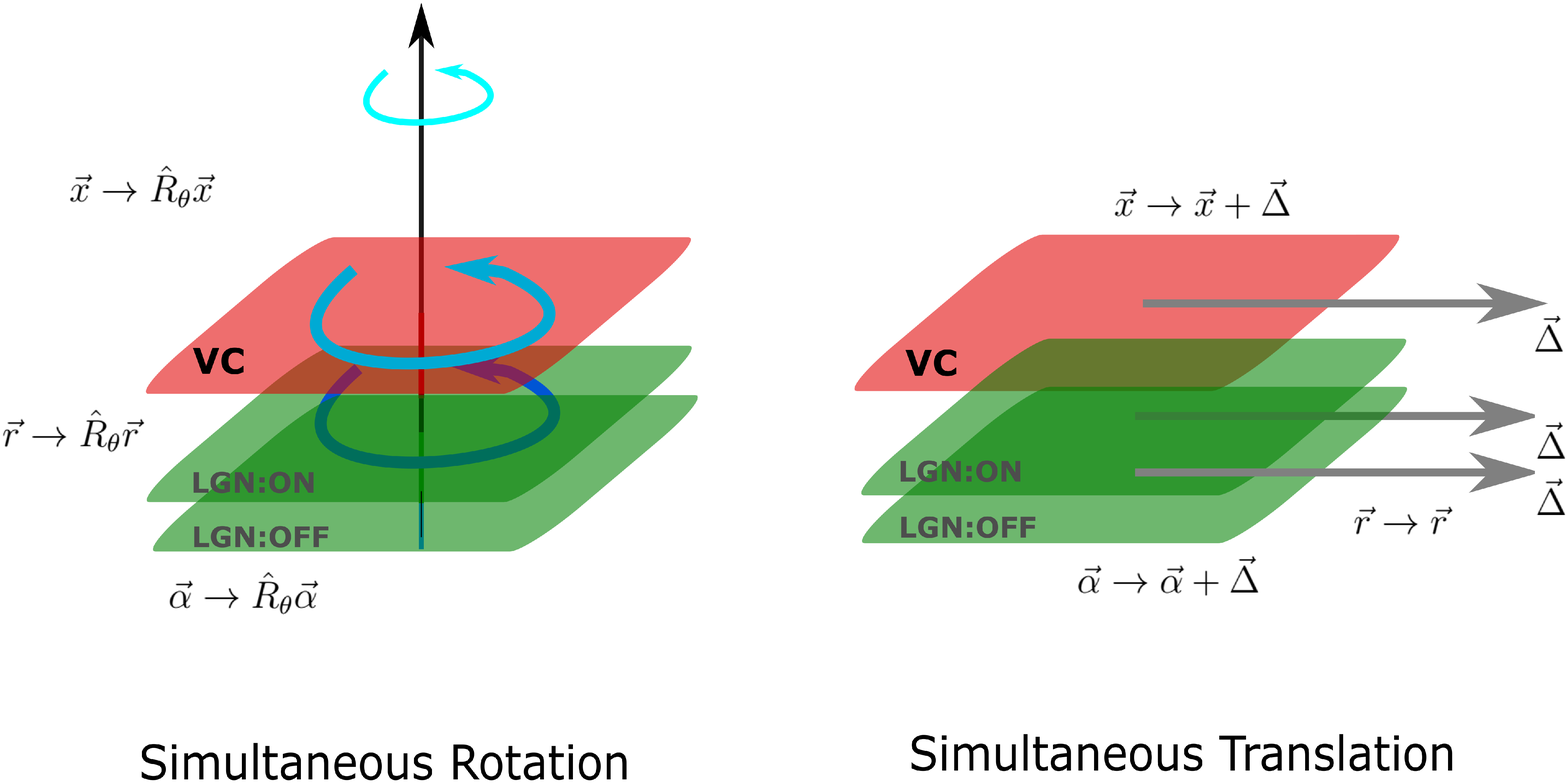}
\centering
\caption{\small{\textbf{Invariances of the theory.}} Depiction of the two main transformation under which the model is invariant: simultaneous rotation and simultaneous translation of the three neuron layers. Notice that the receptive field coordinate $\bm{r}$ is left untransformed by translations.}
\label{fig:invariants}
\end{figure}

Since LGN activity reflects retinal input and we averaged at the outset over an isotropic input ensemble, we expect no change in the dynamics from simultanously rotating both the cortical layer and the two LGN sheets by the same angle. The same is true, as already noted, if we consider simultaneous translations of the three layers (see Fig.~\ref{fig:invariants}). 

The time-evolution operator $\hat{L}^p(\bm{\omega})$ has thus two symmetries: (1) Simultaneous shifts of the cortex and of LGN do not affect the matrix elements; (2) If the wavevector $\bm{\omega}$ is rotated, and the relative coordinates $(\bm{r}, \bm{s})$ are rotated by the same angle, the matrix elements are also unchanged. If we consider the null wavenumber $\omega=0$, the latter operation reduces to rotating the $\bm{r}$-coordinates only, which therefore does not affect the matrix elements.  Since the time-evolution operator at zero wavenumber can be diagonalized simultaneously with the rotation operator, we will occasionally follow McKay and Miller (Neural Computation 2, 2: 169-182, 1990) in referring to its eigenstates through the language of atomic orbitals (1s: nodeless; 2s: one radial node; 2p: one angular node; etc.). 
  
From the discussion of Sec.~\ref{sec:long-term-dyn}, it follows that 
the problem can be treated analogously to the study of zero-temperature phase transitions, in which different phases are often entirely characterized by changes in symmetry. 

Translation symmetry is broken if the principal eigenstate of the system corresponds to a nonzero wavenumber. Rotation symmetry is broken if the principal eigenstate is not invariant under simultaneous rotations of the wavevector and of the radial coordinate. For instance, this happens if the wavenumber under consideration is $\omega =0$ and the eigenfunction has angular momentum $l=1$ (a "p-wave") or any other angle-dependent -- hence orientation-selective -- functional form.
  
It follows that there are multiple symmetry classes for the solution, and it is convenient to introduce shorthand labels for the phases that will emerge from the analysis.  We will call "N-phase" (non-selective) the phase in which no invariance is broken, "R-phase" the phase where rotation symmetry is broken but translation symmetry is not, and "T-phase" the phase where translation symmetry is broken, and so is rotation symmetry. A summary of these phases was given in Table~(\ref{table:phases}). 
    
\subsection{Symmetries of the system: parity and CP symmetry} 
\label{sec:pc}

An important property of the eigenfunctions of $\hat{L}^p$ 
concerns their behavior under the action of the operators 
$\PP_x$ and $\PP_y$, defined by 

\be 
 \PP_x \psi(r_x,r_y) = \psi(-r_x,r_y) \hspace{50 pt} \PP_y \psi(r_x, r_y) = \psi(r_x, -r_y).
 \ee

As we take the wavevector $\bm{\omega}$ to be aligned with the x-axis,  the commutation rule $[\hat{L}^p, \PP_y]=0$ is immediately verified
from Eq.~(\ref{Lp_first_real_space_matrix_elements}), hence $\hat{L}^p$ and $\PP_x$  can be diagonalized simultaneously, and the eigenfunctions of $\hat{L}^p$ may be chosen as either symmetric or antisymmetric under inversion of the $r_y$ coordinate. 

On the other hand, the operators $\hat{L}^p$ and $\PP_x$ do not commute,
as can be seen from Eq.~(\ref{Lp_first_real_space_matrix_elements}). 
However $\hat{L}^p$ does commute with the product $\CC \PP_x$, where $\CC$ is the antilinear operator such that $\CC \psi(r) = \psi^*(r)$.  

Writing the complex RF $\psi(r) = u(r) + i v(r)$  as the real-valued vector function $\psi (r) = \left( \begin{array}{c} u(r)  \\ v(r) \end{array} \right)$, we have that 
\be 
\label{CP_matrix}
\CC\PP_x= \left( \begin{array}{cc} 
\PP_x & 0 \\ 0 & - \PP_x
\end{array} \right) 
\hspace{30pt} \CC\PP_y= \left( \begin{array}{cc} 
\PP_y & 0 \\ 0 & - \PP_y
\end{array} \right). 
\ee
which is an Hermitian operator, so that its eigenvalues must be real.  Since $(\CC\PP_x)^2=(\CC\PP_y)^2=1$, it follows that the eigenvalues are $\pm 1$.  
 
In this representation, a generic integral operator $\hat{O}$ takes the matrix form 
$ \hat{O} = \left( \begin{array}{cc} \hat{A} & -  \hat{B} \\ \hat{B}& \hat{A}
 \end{array} \right)$, where the kernels $A(r,s)$ and $B(r,s)$ of $	\hat{A} $ and $\hat{B}$ are the real and imaginary part of the integral kernel $O(r,s)$ of $\hat{O}$. 
Such an operator clearly commutes with multiplications of the wave functions by an arbitrary "gauge factor" $e^{i \theta}$. Indeed, such a gauge transformation  is represented by the rotation of the complex plane 
    
\begin{equation}
\label{gaugechange} 
\hat{R}_{\theta} = \left( \begin{array}{cc} \cos{\theta}  & - \sin\theta \\ \sin\theta & \cos\theta \end{array} \right), 
\end{equation}
and we have $\left[\hat{O}, \hat{R} \right] = 0$. 
 
If $\left[ \hat{O}, \CC\PP \right]=0$ for the parity operator $\PP$ corresponding to a given coordinate $r$, it 
follows that the eigenfunctions $[u(r), w(r)]$ of $\hat{O}$ can be chosen to be eigenvectors of the operator $\CC \PP$, whose eigenvalues we discussed after Eq.~(\ref{CP_matrix}). That is, they can be chosen to obey the constraint 
\be 
 \left( \begin{array}{c} u(-r) \\ - w(-r) \end{array} \right)  = \CC \PP \left(\begin{array}{c} u(r)\\  w(r) \end{array} \right)   = \lambda_{CP} \left(\begin{array}{c} u(r) \\ w(r) \end{array} \right) = 
\pm \left(\begin{array}{c} u(r) \\ w(r) \end{array} \right),
\ee
from which we can see that either $u(r)$ is symmetric and $w(r)$ antisymmetric, or vice versa. In both cases, the symmetric and antisymmetric part of the function are separated by a phase shift of magnitude $\pi$.

Applying this to the constrained time-evolution operator $\hat{L}^p$, we conclude that its eigenfunctions will consist of a component $\psi_S$ that is symmetric in $\hat{P}_x$ and a component $\psi_S$ that is antisymmetric, the two components being separated by a phase shift $\pi$. 

We can thus write 
\be 
\psi (\bm{r}) \propto \psi_S (\bm{r}) \pm i \psi_A (\bm{r}), 
\ee
where $\psi_S$ and $\psi_A$ are real, and $\psi_S$ ($\psi_A$) an even (odd) function in $r_x$. 
 
Notice that the operator $\CC\PP$ does not commute with the gauge operator $\hat{R}_{\theta}$ defined by Eq.~(\ref{gaugechange}). This means that by diagonalizing $\CC\PP$ we have effectively fixed the gauge of the wave functions. Thus, we have shown that it is $\textit{possible}$ to write the eigenfunctions of $\hat{L}^p$ in the form $\psi_S (\bm{r}) + i \psi_A (\bm{r}) $. If we back-transform to real space in the cortical coordinates $\bm{x}$, this means that symmetric and antisymmetric RFs will alternate along the direction of cortical modulation.
  
\subsection{Diagonalization of the unconstrained dynamics}
\label{sec:diag_unconstr} 
 
In Fourier space, the unconstrained two-layer model is given by Eq.~(\ref{form2}), which can be diagonalized exactly. 

Indeed, if we define the basis transformation $\Psi(\bm{r}) = \exp \left( i \frac{\eta^2}{\mu^2} \omega r_x \right) \chi(r_x, r_y)$, it is clear that $\Psi(\bm{r}) $ is an eigenfunction of $\hat{L}$ if and only if $\chi(\bm{r})$ is an eigenfunction with the same eigenvalue of the integral operator with kernel
\be 
\label{Ld}
L^{\chi}(\bm{r},\bm{s}) = 
\exp\left[
-\frac{ \omega^2}{2 \Omega^2} - \frac{r^2 + s^2}{4 \rho^2} 
- \frac{(\bm{r} - \bm{s})^2}{2 \mu^2} \right],
\ee

The full diagonalization of this operator was first accomplished in Cartesian coordinates by  Wimbauer et al. (Network, 9, 4: 449-466, 1998).
They found that the normalized eigenfunctions have the form 
\begin{equation}
\label{carteseigenf}
\chi_{n_x n_y} (\bm{r}) = \left( 2^{\frac{n_x+n_y}{2}} \sqrt{ \pi n_x! n_y! } \ \gamma \right)^{-1} e^{- \frac{r^2}{2 \gamma^2}}H_{n_x} \left( \frac{r_x}{\gamma} \right) H_{n_y} \left( \frac{r_y}{\gamma} \right),
\end{equation}
 with the number $n_x$ and $n_y$ being nonnegative integer and the functions $H_n$  Hermite polynomials, while the corresponding eigenvalues are 
\be \label{cart_spec} \Lambda_{n_x n_y} = 2 \pi \mu^2 e^{ - \frac{\omega^2}{2 \Omega^2}}   \beta^{-n_x - n_y - 1} \ee
 and the two parameters entering these formulas are 
 \begin{eqnarray}
\label{redefinition}
\gamma = \sqrt{2} \rho \left( 1 + \frac{4 \rho^2}{\mu^2}\right)^{- 1/4},
\hspace{30pt} 
\beta = 1 + \frac{\mu^2}{2 \rho^2} + \frac{\mu}{\rho} \sqrt{1 + \frac{\mu^2}{4 \rho^2} },
\end{eqnarray}

The parameter $\gamma$ is  the width of the receptive fields, which quantifies how the arbor radius $\rho$ is renormalized by recurrence and input correlations. For fixed $\rho$, the width $\gamma$ of the eigenfunctions is a monotonically decreasing function of the ratio $\rho/\mu$. In this representation (having divided the RFs by $\sqrt{A(r)}$ at the outset in Sec.~\ref{sec:homeostatic}) the unrenormalized arbor radius is represented by $\sqrt{2} \rho$. It follows that, if the ratio $\rho/\mu$ is very small, no renormalization occurs:  $\gamma = \sqrt{2} \rho$. 
  
If the ratio $\rho/\mu$ tends to infinity (i.e., if the arbors are comparatively wide, asymptotically extending over all the cortex) the range of the eigenfunction will be restricted by the correlation-interaction length scale, becoming equal to the geometric mean of the two length scales; namely, $\gamma \sim \sqrt{\rho \mu}$. 

For future reference, we note the three highest-lying eigenfunctions $\Psi_{n_x, n_y}$ of $\hat{L}$: 
\bea 
\label{Psi00}
\Psi_{0, 0} ( \bm{r}) =\frac{1}{\sqrt{\pi} \gamma} 
\exp \left( i \frac{\eta^2}{\mu^2} \omega r_x - \frac{r^2}{2 \gamma^2}\right); \\
\label{Psi01}
\Psi_{0,1} (\bm{r}) = \frac{r_x}{\sqrt{\pi} \gamma^2}
\exp \left( i \frac{\eta^2}{\mu^2} \omega r_x   - \frac{r^2}{2 \gamma^2}   \right); 
\\
\label{Psi10}
\Psi_{1,0} (\bm{r}) = \frac{r_y}{\sqrt{\pi} \gamma^2}
\exp \left( i \frac{\eta^2}{\mu^2} \omega r_x   - \frac{r^2}{2 \gamma^2}   \right); 
\eea
   
Since the operator $\hat{L}^{\chi}$  is symmetric with respect to rotations of the vector $\bm{r}$, it can also be diagonalized simultaneously with the generator of  rotations for the vector $\bm{r}$, as done more recently by Davey et al. (arXiv:1805.03749). This leads to writing the eigenvalues of $\hat{L}$ in the equivalent angular form 
\begin{equation}
\label{Linsker_spectrum}
\lambda_{N,m} = 2 \pi \mu^2
 e^{ - \frac{\omega^2}{2 \Omega^2}}   \beta^{-2 N - m - 1},
\end{equation}
where the integer $m$ is the angular momentum, or the number of angular nodes in the eigenfunctions,  of $\hat{L}^\chi$ while $N$ is their number of radial nodes, and we defined
$$\beta \equiv 1/q = 1 + \frac{\eta^2}{2 \rho^2} + \frac{\eta}{\rho} \sqrt{1 + \frac{\eta^2}{4 \rho^2} }.$$ 

The corresponding eigenfunctions $\Phi_{N,m}$ of $\hat{L}$ are best written as functions of polar coordinates ($r, \phi$). The highest such eigenfunctions are 
\bea 
\label{0_0_eigenf}
\Phi_{0, 0} ( \bm{r}) &=& \Psi_{0,0} (\bm{r}) 
\\
\label{0_1_eigenf}
\Phi_{0,1}^{\pm} (\bm{r}) &=& \frac{r}{\sqrt{\pi} \gamma^2} \exp\left(  
i \frac{\eta^2}{\mu^2} \omega r \cos\phi
- \frac{r^2}{2 \gamma^2}  \pm i \phi \right); 
\\
\label{1_0_eigenf}
\Phi_{1, 0} ( \bm{r}) &=&
\frac{(\gamma^2 - r^2) }{\sqrt{ \pi } \gamma^3 } \exp\left(i \frac{\eta^2}{\mu^2} \omega r \cos \phi- \frac{r^2}{2 \gamma^2}\right);
\eea 

Eigenfunctions of the angular-momentum representation with an even (odd) number of angular nodes are built with appropriate Clebsch–Gordan coefficients from eigenfunctions of the Cartesian representation where $n_x$ and $n_y$ have same (different) parity.

From formulas (\ref{cart_spec}, \ref{Linsker_spectrum}), it is seen that the dependence of the eigenvalue on the wavenumber  lies entirely in the exponential prefactor. Hence, the optimal  wavenumber is always $\omega=0$. Translation symmetry is never broken in the absence of homeostatic constraints.
  
Since $\hat{L}$ is diagonalizable, and the other operators summed into $\hat{L}^p$ have separable matrix elements, $\textit{each}$  of the four operators summing up to $\hat{L}^p$ in Eq.~(\ref{form1}) is diagonalizable exactly. Unfortunately the sum of the four is not. But while no closed-form solution is available in general, it will be possible to study the operator separately in various regions of parameter space. 
 
\section{Cortically uniform phases} 
\label{sec:uniform}

Let us assume that, for some given values of $\zeta$ and $\eta$,  the principal eigenfunction has the form 
$\Psi (\bm{x},\bm{r}) \equiv \Psi(\bm{r})$, which is uniform over cortex, or in other words
 that the principle eigenfunction in that point of parameter space is $\omega=0$ . We will reform to such regions as "uniform phases". We would like to know, given a point in parameter space where such a phase is dominant, whether it will be of the R or N type. 
  
It can be seen that the operator $\hat{L}^p$ of Eq.~(\ref{form1}), acting in such a case on functions of the single variable $\bm{r}$, becomes equal to the operator $\hat{L}^f$ with matrix elements
\begin{eqnarray}\nonumber
\frac{ L^f(\bm{r}, \bm{s}) }{ \sqrt{A(r)  A(s)} } = I^\mu  \left( \bm{r},\bm{s} \right) 
+
\int d\bm{r}_1 d \bm{s}_1 A(r_1) A(s_1) I^\mu  \left( \bm{r}_1 , \bm{s}_1\right) \\ -  \int d\bm{u} A(u) I\mu  \left( \bm{r} , \bm{u} \right)
 -  \int d\bm{u} A(u) I\mu  \left( \bm{u} , \bm{s} \right) 
\label{Lf_to_integrate}
\end{eqnarray}

where $ I^\mu  \equiv  = \exp\left[ - \frac{(\bm{x}-\bm{y})^2}{2 \mu^2}\right]$ is a version of the interaction function in (\ref{gaussian_choice_interaction}) 
corrected by the input. 

We can also rewrite  Eq.~(\ref{Lf_to_integrate}) compactly as:
\begin{equation}
\label{Lf_operator}
\hat{L}^f = \hat{L}^{\mu} + | a_0 \rangle \langle a_0 | \hat{L 
}^{\mu} | a_0 \rangle \langle a_0 | - 2 \ \text{HP} \left[ \hat{L 
}^{\mu}| a_0 \rangle \langle a_0 |   \right],
\end{equation}
where "HP" is the Hermitian part of an operator and the unconstrained part of the $\hat{L}^f$ operator has matrix elements 

\be
\label{L_mu_matrix_el}
L^{\mu}(\bm{r},\bm{s}; \mu) = \sqrt{A(r) A(s) } I^\mu(\bm{r},\bm{s})
\ee

The  rest of this section is devoted to the diagonalization of $\hat{L}^f$, which we will perform by treating separately the regimes with large and small values of $\mu/\rho$.  

\subsection{Long effective length ($\mu \gg \rho$)}
\label{long-eff-length}

Although the operator $\hat{L}^f$  is not amenable to exact diagonalization, it is easy to show that, in the regime of long effective length  ($\mu  \gg \rho$), rotation symmetry is broken, leading to the development of orientation selectivity. 
 
To see this, assume self-consistently that all the radial variables in the eigenvalue equation for $\hat{L}^f$ will be confined to a region of order $\rho$.  Hence, expanding the unconstrained operator in Eq.~(\ref{L0_matrix_el}) can be expanded as 
\be 
\label{L0_expanded}
L^{\mu} (\bm{r},\bm{s} ; \mu)  \approx e^{ - \frac{r^2 + s^2}{4 \rho^2}} \left[ 1 - \frac{(\bm{r} - \bm{s})^2}{2 \mu^2}  \right] \ee 
where further corrections inside the square brackets are of order $\left( \frac{\rho}{\mu} \right)^4 $  
 
If we substitute (\ref{L0_expanded}) into (\ref{LcExpression}), we find that in the asymptotic matrix element the terms of order $\left( \frac{\rho}{\mu} \right)^0 $ vanish exactly. The terms of order $\left( \frac{\rho}{\mu} \right)^2 $ cancel each other leaving only the following: 
\be 
\label{longrangelcmatrixelement}
L^f(\bm{r,s}; \mu) \approx \frac{ \bm{r \cdot s}}{\mu^2} \sqrt{A(r) A(s)}  = \frac{ r s }{\mu^2}  \sqrt{A(r) A(s)}   \cos(\phi_r - \phi_s)
\ee 
while further corrections are again of order $\left( \frac{\rho}{\mu} \right)^4 $.

It is clear that the only positive eigenvalue of the operator defined by the kernel (\ref{longrangelcmatrixelement}) corresponds to the eigenfunction
\be 
\label{longrange_eigenf}
\psi (\bm{r}) \propto r \sqrt{A(r)} \cos(\phi - \phi_0),
\ee
the corresponding eigenvalue being just the p-wave eigenvalue $\lambda_{0,1}$ of the unconstrained model. 
  
The expansion is self-consistent because indeed the function (\ref{longrange_eigenf}) vanishes for $r \gg \rho$.  

Following a convention in the literature (MacKay and Miller, Neural Computation, 2,2:169-182, 1990), we will refer to eigenfunctions $\psi(\bm{r}) = f(r)$ as "s-wave" states. We will call "p-wave" states eigenfunctions having angular momentum $m=1$, i.e. with angular dependence $\cos \left(m(\phi-\phi_0)\right)$ with $m=1$. From Eq.~(\ref{longrangelcmatrixelement}), we see that all s-wave eigenstates have zero eigenvalue to this order in the expansion. For sufficiently long effective length $\mu$, the principal eigenspace is thus composed by the p-wave functions described in Eq.~(\ref{longrange_eigenf}).

\subsection{Short  effective length ($\mu \ll \rho$)} 
\label{short-eff-length}

We have argued that p-waves dominate the uniform phases in the limit of long effective length $\mu \gg \rho$. We would like to inquire whether there exist regions of parameters where this is not the case, i.e. where rotation symmetry is not broken and s-waves dominate the uniform phases. These s-waves would describe RFs that are unable to discriminate among the possible orientations of visual input. 

If that is the case,  there can be no smooth crossover between the two regimes.  
A linear combination of an s-wave $(m=0)$ and of a p-wave $(m=1)$ could not be an eigenfunction of $\hat{L}^f$ other than at special points of degeneracy. Let us tentatively call  $\Theta_c= \mu_c/ \rho$ the largest value of $h=\mu/\rho$ where the principal eigenfunction is non-selective. We would like to find if $\Theta_c >0$ and, if so, compute the structure of the receptive field for $h<\Theta_c$.  

A natural tool to address this question is the variational method for linear operators. We assume a functional form (trial function) for the principal eigenfunction; we normalize it; we find the expectation value of our operator in that state; and we maximize it with respect to variational parameters. This leads to the best available approximation of the principal eigenvalue within the given Hilbert subspace. 

The expectation value of the operator $\hat{L}^f$ in the trial state  $| \psi  \rangle$ is defined as 
\begin{equation}
\label{exp_value}
E [\psi] = \frac{ \langle \psi | \hat{L}^f |\psi \rangle }{\langle \psi | \psi \rangle}. 
\end{equation}

It can be  shown by the same arguments as in MacKay and Miller (Neural Computation 2, 2: 169-182, 1990) that the principal eigenfunction of $\hat{L}^f$ in the s-sector must be of the 2s type, i.e. with one radial node. We will thus choose our variational trial function to be a RF with the same functional form as the 2s eigenfunction of the unconstrained model, only with the position of the node unspecified. 
    
The unconstrained 2s wave function is, as per Eq.~(\ref{1_0_eigenf}), a Gaussian RF of width $\gamma$ multiplied by the polynomial $(\gamma^2 - r^2)$, so that  the radial node is located at $r = \gamma$. We will now replace the nodal radius $\gamma$ with an unspecified radius $R$, obtaining a trial function that is a generalization of Eq.~(\ref{1_0_eigenf}), and will optimize the expectation value of $\hat{L}^f$ with respect to $R$ over all Hilbert space. Our "movable-node" trial function is therefore 
\be  
\label{moveable_node}
\psi_T^{(R)}(\bm{r}) = \frac{N}{\sqrt{\pi} \gamma} (R^2 - r^2) \exp\left( - \frac{r^2}{2 \gamma^2} \right), 
\ee 
where the value of $\gamma$ is given by Eq.~(\ref{redefinition}) and we have introduced the normalization factor $N = \left[ \gamma^4 + \left( \gamma^2 - R^2 \right)^2 \right]^{-1/2}$. 
  
Let us consider the expectation value Eq.~(\ref{exp_value}) of the unconstrained operator of Eq.~(\ref{L0_matrix_el}) in the state~(\ref{moveable_node}). This is given by 
\begin{eqnarray}
&& \mathcal{E}_0 (R) \equiv \langle \psi_T^{(R)} | \hat{L}^\mu | \psi_T^{(R)} \rangle
\\ 
\nonumber 
&=& \frac{N^2}{\pi \gamma^2} \times 4 \pi^2 \int_{0}^{\infty} dr \ r (R^2 - r^2) \int_0^{\infty} \ ds  \ s \ (R^2 - s^2)
I_0\left(\frac{ rs }{\mu^2}\right) e^{- \left( \frac{1}{ 2 \rho^2} + \frac{1}{\mu^2} + \frac{1}{\gamma^2} \right) \frac{r^2 + s^2}{2}}.
\label{E_0big}
\end{eqnarray}
which, after integration, yields 
\be  \mathcal{E}_0 (R) = \frac{2 \pi \mu^2}{\beta^3} \ \frac{1 + \beta^2 \left( 1- R^2/\gamma^2\right)^2}{1 + \left(1 - R^2/\gamma^2\right)^2 }.
\ee
with $\beta$ defined in Eq.~(\ref{redefinition}).

To optimize this expectation value, we need to maximize  the function $f(x) = \frac{1 + \beta^2 x}{1 + x}$, where $x= (R^2/\gamma^2 - 1)^2$. The derivative is $f'(x) = \frac{\beta^2 - 1}{(1 + x)^2} $ , always nonnegative because $\beta \geq 1$; hence it will be sufficient to maximize $x$, which is done by choosing the limit $R \rightarrow \infty$. The result is unsurprising: in the limit $R\rightarrow \infty$, the moveable-node function becomes in fact nodeless, and it is nothing but the 1s Gaussian of width $\gamma$ which we know as the principal eigenfunction of $\hat{L}^\mu$. 
    
Let us now consider the expectation value Eq.~(\ref{exp_value}) of the full operator $\hat{L}^f$, as described by Eq.~(\ref{Lc_operator}), calculated in the moveable-node state of Eq.~(\ref{moveable_node}). This can be written as 
\be 
\label{somewhat_intricate}
 \mathcal{E} (R) = \mathcal{E}_0 (R) + \frac{\mu^2 A^2}{\mu^2 + 2 \rho^2} - \frac{2 \mu^2 AB}{\mu^2 + \rho^2}, 
 \ee
where 
\begin{eqnarray} A&=& \frac{N}{\sqrt{\pi} \gamma} \times 2 \pi \int_0^{\infty} (R^2 - r^2) \exp\left[ - \left( \frac{1}{\gamma^2} + \frac{1}{2 \rho^2} \right) \frac{r^2}{2} \right] r \ d r 
\\ 
B&=& \frac{N}{\sqrt{\pi} \gamma} \times 2 \pi \int_0^{\infty} (R^2 - r^2) \exp\left[ - \left( \frac{1}{\gamma^2} + \frac{1}{2 \rho^2} + \frac{1}{\rho^2 + \mu^2 }\right) \frac{r^2}{2}  \right] r \ d r 
\end{eqnarray}
or, upon integration, 
\begin{eqnarray}
\label{A_intricate}
A &=& 4 \sqrt{\pi} N \gamma \rho^2 \frac{(2 \rho^2 + \gamma^2) R^2 - 4 \gamma^2 \rho^2}{(2 \rho^2 + \gamma^2)^2}
\\
\label{B_intricate}
B &=& 4 \sqrt{\pi} N \gamma \rho^2 (\rho^2 + \mu^2) \frac{2 \rho^2 ( R^2 - 2 \gamma^2) (\rho^2 + \mu^2) + R^2 \gamma^2 (3 \rho^2 + \mu^2) }{\left[ (\rho^2 + \mu^2) (2 \rho^2 + \gamma^2) + 2 \gamma^2 \rho^2 \right]^2}.
\end{eqnarray}
 
\begin{figure}[htp]
\includegraphics[width=.7\textwidth]{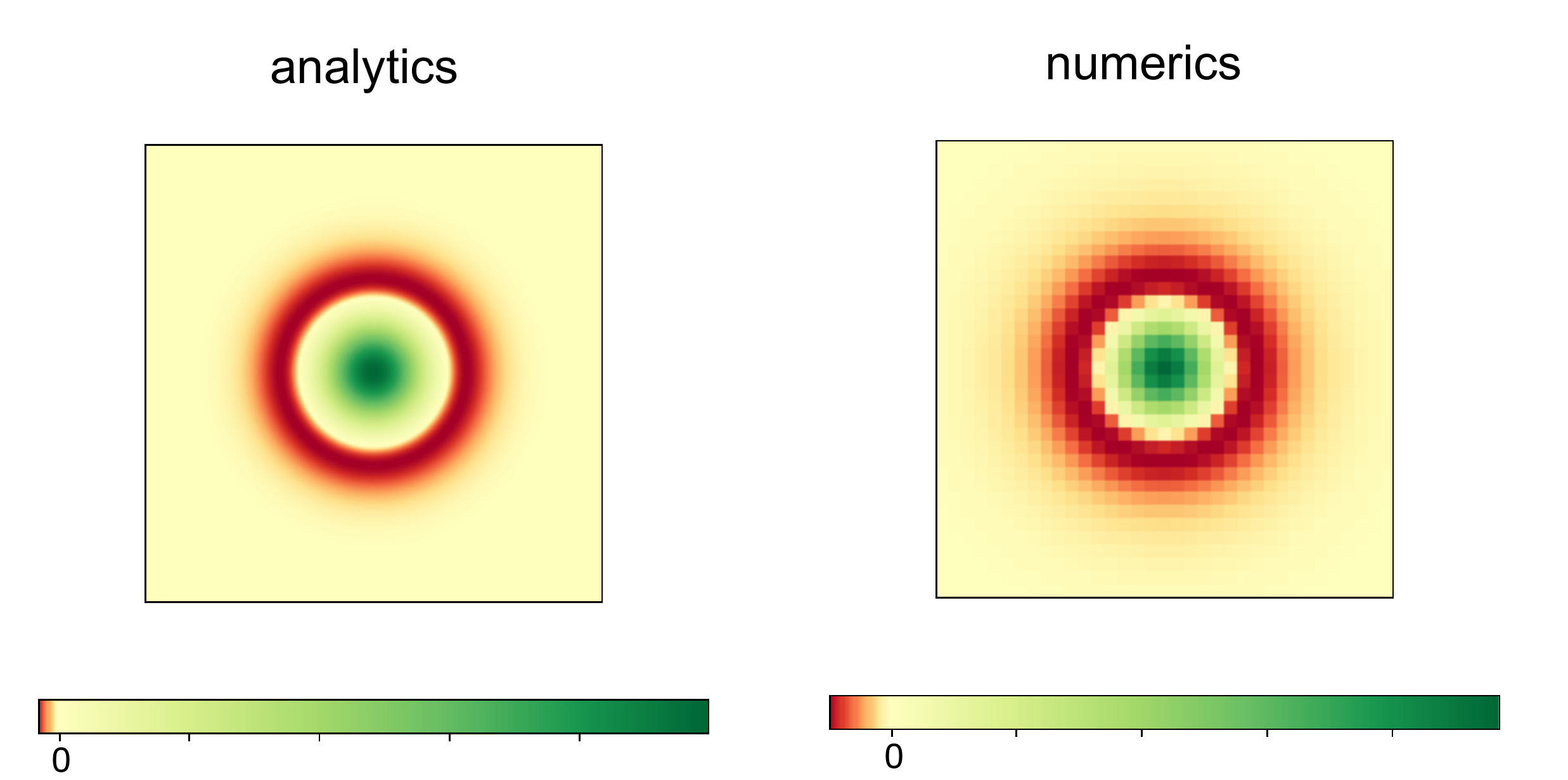}
\centering    
\caption{
\small{\textbf{Receptive fields for the N-phase.}}
Example of the variational RF of  formula~\ref{psiN}, compared to the result of numerically diagonalizing the full operator $\hat{L}^p$ and rescaling the eigenfunction by Eq.~\ref{t_to_s}. The parameters used here are $\zeta/\rho=0.02$,$\eta\rho = 0.2$.A  side of the grid has length equal to $5 \rho$, color scale ranges between min and values.}
\label{fig:N_check}
\end{figure}

While expression (\ref{somewhat_intricate}) with the substitution of (\ref{A_intricate}-\ref{B_intricate}) 
is somewhat intricate, we are ultimately interested only in its maximal value over all the range of nodal radii $R$. We thus expand $\mathcal{E}$ in $h=\mu/\rho$ with the ansatz $R^2 = \rho^2 \left( k^2  h + O(h^2)\right) $, yielding 
\be 
\frac{\mathcal{E}(\rho k \sqrt{h})}{2 \pi \mu^2} = 1 - 2 f(k)  h + O(h^2),
\ee
where $f(k) = \frac{ 8 - 7 k^2  + 2 k^4}{2 - 2 k^2 + k^4}$.  The requirements $f'(\bar{k}) = 0, f''(\bar{k}) >0$ lead to 
\be 
\label{kbar_eq}
\bar{k} = \sqrt{\frac{4 + \sqrt{10}}{3}},
\ee 
which means that the node behaves as $R \sim \left( \frac{4 + \sqrt{10}}{3}  \mu \rho \right)^{1/2}$. 
 Inserting this into Eq.~\ref{moveable_node} and applying 
Eq.~\ref{t_to_s} leads straight to formula~\ref{psiN} of the main text (for a  comparison with numerics see Fig.~\ref{fig:N_check}).
 
Further substituting into Eq.~\ref{somewhat_intricate}, we find that the optimal expectation value is 
\be 
\label{kbar_E}
\mathcal{E}  \equiv \mathcal{E} (\bar{k} \sqrt{h}) =  2 \pi \mu^2 \left[ 1 - (5 - \sqrt{10}) h \right].
\ee
 
We can now compare $\mathcal{E}$ with the exact eigenvalue of the dominant p-wave,  which is given by Eq.~(\ref{Linsker_spectrum}) as $\lambda_{0,1} = 2 \pi \mu^2 / \beta^2 \sim 1 - 2 h$. Since  $(5 - \sqrt{10}) \sim 1.83  < 2$, we conclude that the principal s-wave eigenvalue approximated by Eq.~(\ref{moveable_node}) lies \textit{higher}. Therefore, the s-waves do indeed dominate for small $h= \frac{\mu }{ \rho}$. 

\begin{figure}[htp]
\includegraphics[width=.5\textwidth]{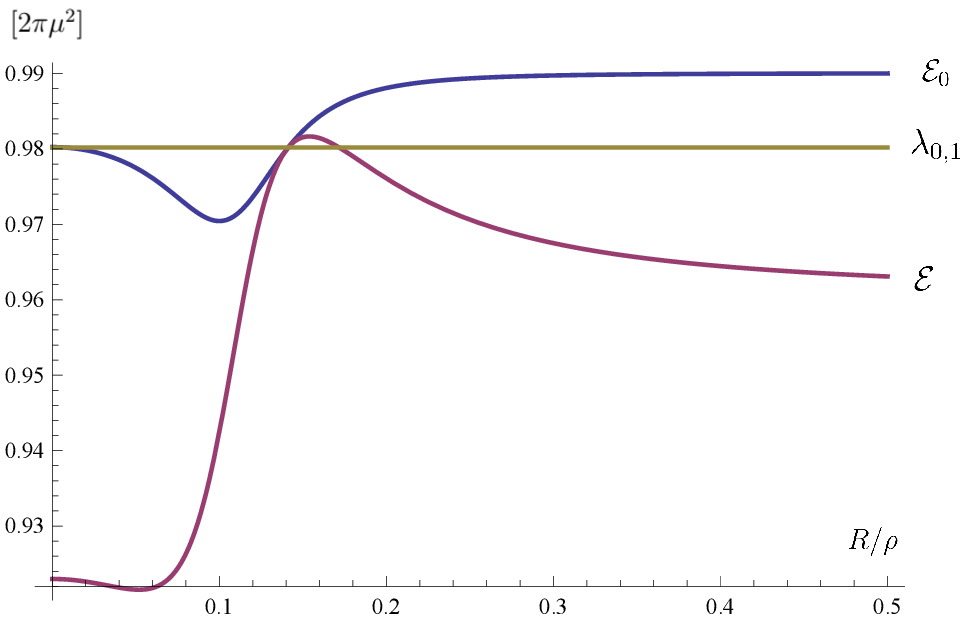}
\centering   
\caption{
\small{\textbf{Variational landscape for cortically uniform phases.}}
Expectation values of the constrained time-evolution operator $\hat{L}^f$ (in units of $2 \pi \mu^2$) plotted as a function of $R/\rho$.  The three curves refer to: 
(1) $\lambda_{0,1}$ (expectation value of $\hat{L}^\mu$  or $\hat{L}^f$ in the exact eigenfunction $\Psi_{0,1}$), plotted in green; (2) $\mathcal{E}_0$, expectation value of the unconstrained operator $\hat{L}^\mu$ in the moveable-node state $\psi_T$, plotted in  blue; (3) $\mathcal{E}$, expectation value of the constrained operator $\hat{L}^f$ in the state $\psi_T$, plotted in purple. The figure refers to $h=\frac{\mu}{\rho} = .01$. Values of $R$ for which the moveable-node state is preferred to the orientation-selective state are different for the two operators
$\hat{L}^\mu$ and $\hat{L}^f$.  Namely, there exists a minimal value $\tilde{R}$, in this example being $\approx 0.14$, such that  $\hat{L}^\mu$ opts for $\psi_T$ at sufficiently high values of the node radius, $R> \tilde{R}$, while $\hat{L}^f$ does so for values of $R$ in a narrow window $R \gtrsim \tilde{R}$. }
\label{fig:moveable_node}
\end{figure}

The eigenvalue landscape leading to dominance of s-waves is displayed in full in Fig.~\ref{fig:moveable_node} for a fixed (sufficiently low) value of  $h=\frac{\mu}{\rho}$. As can be seen, choices of the movable node below a certain threshold $\tilde{R}(h)$ would lead to s dominance, but the optimal  $R$ in the presence of constraints (purple curve) lies beyond that threshold.
  
\subsection{Phase boundary of the uniform phases} 
\label{sec:uniform_boundary}

We would like now to have a lower bound on the critical value of the interaction length $\mu_c = \Theta_c\rho$ at which rotation symmetry is first broken. We can define $\Theta_c$ as the largest value of $\mu/\rho$ where the s-mode dominates. We proceed by finding the $h$ for which the expectation value of Eq.~(\ref{kbar_E}) is equal to the exact $2p$ eigenvalue. The fact that this will indeed yield a lower bound on the actual value of the transition point can be understood as follows.  

If the variational method reveals the transition at a point $h= \theta_c$, it means that we have found an s-wave state whose expectation value is larger than the exact eigenvalue of the principal p-wave for all $h < \theta_c$. Suppose ad absurdum that the actual critical point $\Theta_c$ is $\Theta_c < \theta_c$. That means in the region $\Theta_c< h < \theta_c$ the actual principal state of the operator is a orientation-selective, i.e. $m>0$. And since the $m>0$ sector is exactly diagonalizable, this principal p-wave  must be the one we already calculated, with eigenvalue $\lambda_{0,1}$. 
 
But if that was true, all the s-wave functions would yield expectation values lower than that eigenvalue. Then it would not be possible to create a linear combination of them (our trial function) that yields an expectation value $> \lambda_{0,1}$, as we have done. We deduce that we must have $\Theta_c \geq \theta_c$. That is, the variational method provides a lower bound on the actual critical point. 
    
Let us proceed with the calculation. We first expand $R$ to  a higher order, as $ R^2/\rho^2 = c_1 h + c_2 h^2 + O(h^3)$. The coefficient $c_1$ can be determined by maximizing Eq.~(\ref{somewhat_intricate}) to the order $O(h)$, which gives $c_1 = \bar{k}^2 = \frac{4 + \sqrt{10}}{3}$. Then we calculate the second term in the expansion of $\mathcal{E} $, plug in the value of $c_1$ we found, and maximize with respect to $c_2$. This second-order correction, computed at the optimal value of $c_2$, is then included in the expectation value, and the whole thing is compared to the eigenvalue of the leading p-waves, to see which is dominating. 
One obtains
\be 
\frac{\mathcal{E}}{2 \pi \mu^2} \sim 1 - \left( 5 - \sqrt{10}\right) h + \left( \frac{33}{2} - \frac{51}{\sqrt{10}} \right) h^2. \label{lambdaMN} 
\ee
 
The critical point $\theta_c$ is found where this s-wave expectation value intersects the p-wave eigenvalue given by Eq.~(\ref{Linsker_spectrum}), that is, 
\be 
 \frac{\lambda_{0,1}}{2 \pi \mu^2} \sim 1 - 2 h + 2 h^2;
 \ee
setting $\mathcal{E} = \lambda_{0,1}$  yields
\begin{equation}
\label{theta_c}
\theta_c= \frac{2 (75 - 8 \sqrt{10} )}{997} \approx 0.1,
\end{equation}
which is a rigorous lower bound to the critical point. The actual value is easiest to find numerically by projecting the operator $\hat{L}^f$ into the $m=0$ subspace and thus turned into an operator $L^{(s)}$ acting on functions of the sole radial variable, whose matrix element is 
\begin{eqnarray}
\hspace{-30pt}  \nonumber L^{(s)}(r, s)  &=& I_0\left(\frac{r s }{\mu^2}\right) e^{- \left( \frac{1}{2 \rho^2} +\frac{1}{\mu^2}\right) \frac{r^2 + s^2}{2} } 
\\ &+&  \frac{\mu^2 }{\mu^2 + 2 \rho^2} e^{ -\frac{r^2 + s^2}{4 \rho^2} }- \frac{\mu^2}{ \mu^2 + \rho^2} \left( e^{ - \frac{r^2}{4  \tilde{\rho}^2}  - \frac{s^2}{4 \rho^2}} + e^{ -\frac{r^2}{4 \rho^2} - \frac{s^2}{4\tilde{\rho} ^2}}\right),
\label{Ls}
\end{eqnarray}
where $ \tilde{\rho} = \left( \frac{1}{ \rho^2} + \frac{2}{\mu^2 + \rho^2 } \right)^{-1/2}$
and $I_0$ is the modified Bessel function of the first kind.  The principal eigenvalue of $L^{(s)}$ must be compared to the exact p-wave eigenvalue so as to obtain the transition point, yielding $\Theta_c \approx 0.34$. 
 
\begin{figure}[htp]
\includegraphics[width=0.6\textwidth]{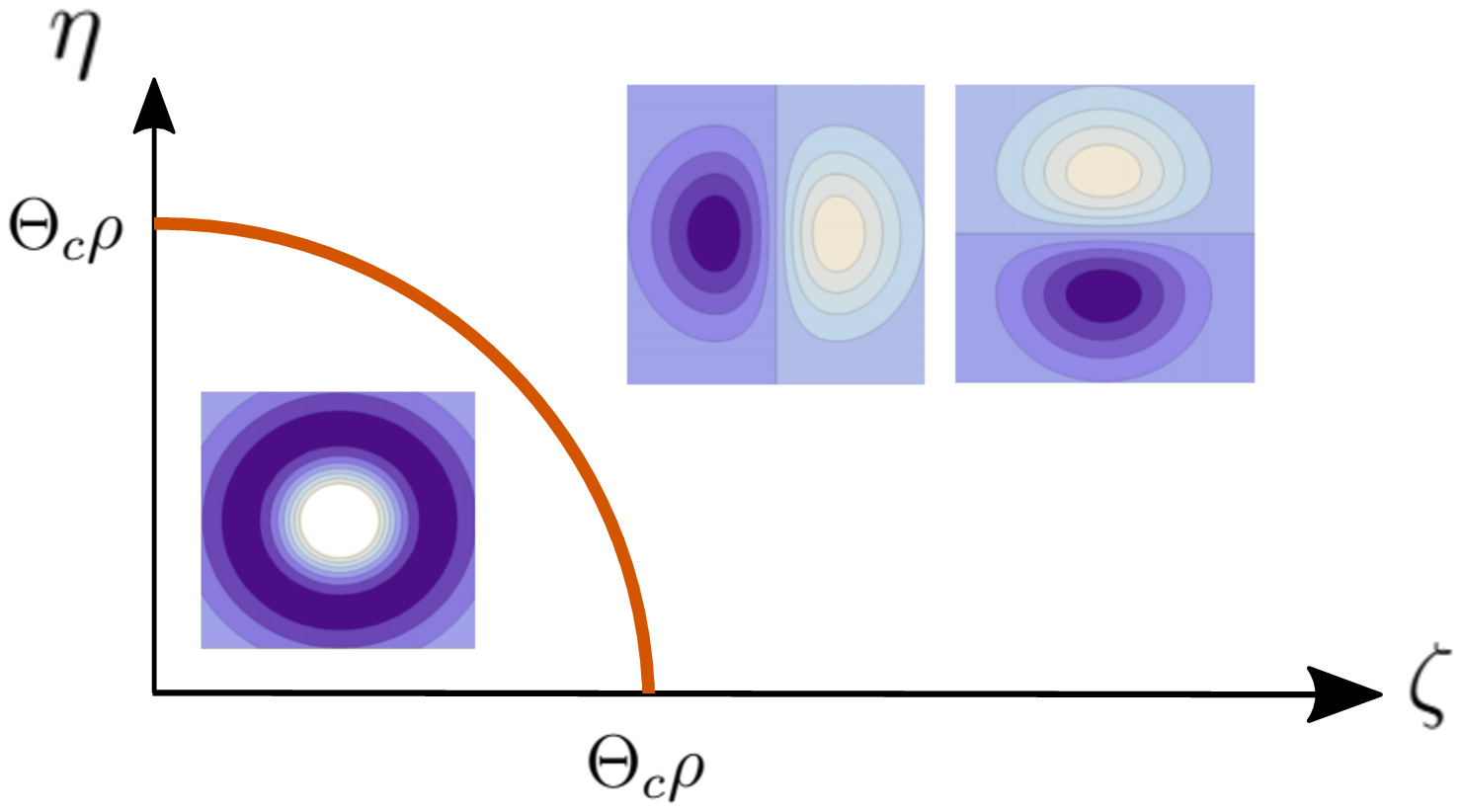}
\centering
\caption{\small{\textbf{structure of cortically uniform phases over the phase diagram.}} Insets show plots of the principal eigenfunctions, respectively the  movable-node approximation for the N-phase and the exact eigenfunctions with longitudinal and 
transverse alignments for the R-phase.} 
\label{fig:uniform}
\end{figure}

The resulting phase diagram  for uniform phases is illustrated in Fig.~\ref{fig:uniform}. In terms of the general model (which allows modulation across cortex as well) the regions within and outside the quarter-circle of Fig. \ref{fig:uniform} can be taken to identify forbidden regions for the R and N phase. Within the quarter-circle ($\mu<\Theta_c \rho$) the R-phase is forbidden because, if at any point the optimal wavenumber happens to be zero, it must yield an N-phase instead. Outside the quarter-circle ($\mu > \Theta_c \rho$) the N-phase is forbidden because, if the optimal wavenumber is zero, it must yield an R-phase instead. 

\subsection{Degeneracies in the uniform phases} 
\label{sec:degeneracies}

To sum up, we have found that if the uniform phase is dominant
 ($\omega  =0$) , the principal eigenfunction of the operator is orientation selective if $\mu > \Theta_c \rho $ and non-selective if $\mu < \Theta_c \rho $, with the critical ratio $\Theta_c $ bounded from below by the value $\theta_c$ of Eq.~(\ref{theta_c}). 
   
For  $\mu > \Theta_c \rho $, there is a two-dimensional degeneracy in the orientation of the symmetry breaking. The exact principal eigenfunction as 
per Eqs.~(\ref{0_1_eigenf},\ref{1_0_eigenf}) is given by
\be  \Psi(\bm{x}, \bm{r})= \left(k_+e^{+i\phi} + k_- e^{-i\phi}\right) \exp\left[ - \frac{r^2}{2 \gamma^2} \right], 
\ee
for arbitrary coefficients $k_+$ and $k_-$.  We will refer in particular to the combinations 
\begin{equation}
\label{psi_xy}
\left( \begin{array}{c}
\Psi^{x}(\bm{x}, \bm{r}) 
\\
\Psi^{y}(\bm{x}, \bm{r}) \end{array} \right) = 
\left( \begin{array}{c}
r_x \\
r_y
\end{array} \right)  \exp\left(  -\frac{r_x^2 + r_y^2}{2 \gamma^2}
\right)
\end{equation}
which are the instances of the 2p waves $\Phi_{0,1}$ and $\Phi_{1,0}$ corresponding to uncorrelated input. 

\begin{figure}[htp]
\includegraphics[width=.7\textwidth]{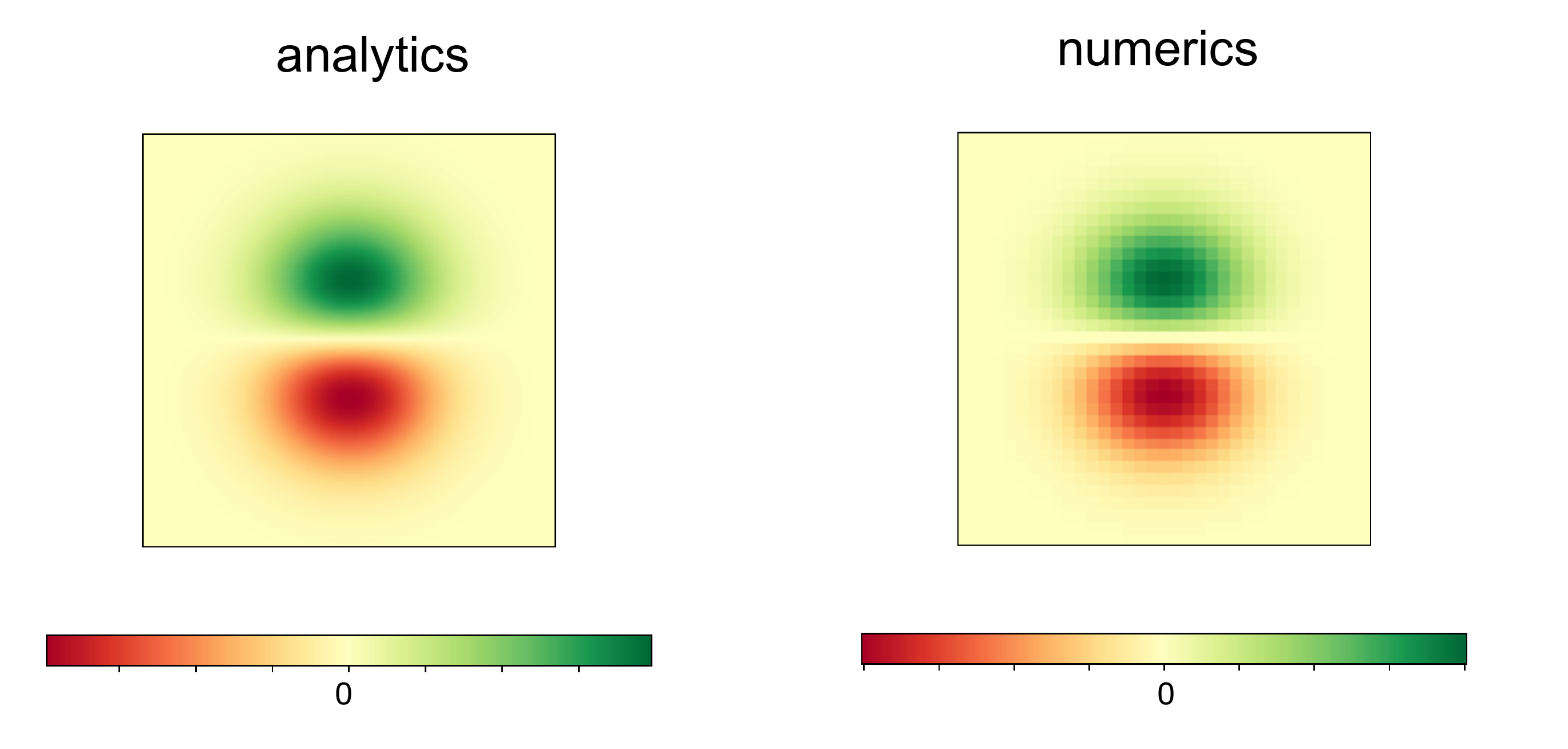}
\centering    
\caption{\small{\textbf{Receptive fields for the R-phase.}}
Example of the  RF of  formula~\ref{psiR}, compared to the result of numerically diagonalizing the full operator $\hat{L}^p$ and rescaling the eigenfunction by Eq.~\ref{t_to_s}. The parameters used here are $\zeta/\rho=0.05$,$\eta\rho = 0.7$. A side of the grid has length equal to $5 \rho$, color scale ranges between min and values.}
\label{fig:R_check}
\end{figure}

Since the cortical wavevector is aligned along the $x$-axis, $\Psi^x$ describes RFs aligned parallel to the cortical wavevector, and $\Psi^y$ describes RFs aligned orthogonally to it. Accordingly, we will call $\Psi^x$ the longitudinal eigenfunction and $\Psi^y$ the transverse eigenfunction. Formula~\ref{psiR} of the main text is obtained from Eq.~\ref{psi_xy} by applying 
 Eq.~\ref{t_to_s}, and is compared to numerics in Fig.~\ref{fig:R_check}.

These two eigenfunctions share, as per Eq.~(\ref{Linsker_spectrum}), 
the exact eigenvalue
\be 
\Lambda^{x,y} =   2 \pi \mu^2 \left( 1 + \frac{\mu^2}{2 \rho^2} + \frac{\mu}{\rho} \sqrt{1 + \frac{\mu^2}{4 \rho^2} } 
\right)^{-2};
\ee

\section{The long-range limit }
\label{sec:long_range}

\subsection{The long-range limit: derivation}
\label{sec:long-range-derivation}
 
We call the long-range limit ($\mu \gg \rho$) the case where either cortical interactions are long-range ($\eta \gg \rho$) or LGN interactions are ($\zeta \gg \rho$) or both. 

We begin by Taylor expanding Eq.~(\ref{form2}) into 
\be
\label{long_range_Lp}
\hat{L}^p \sim \hat{L}_0 + \hat{R} + \hat{S}  + \hat{T} + \hat{T}^{\dagger},
\ee
with 
\begin{eqnarray}
\label{L0_again}
L_0 
(\bm{r,s}; \bm{\omega}) &=&\exp\left[
-\frac{ \omega^2}{2 \Omega^2} - i \frac{\eta^2}{\mu^2} \bm{\omega} (\bm{r} - \bm{s}) - \frac{r^2 + s^2}{4 \hat{\rho}^2} \right],
\\
R
(\bm{r,s}; \bm{\omega}) &=& \left(\frac{r_i s_i}{\mu^2} 
+ \frac{r_i r_j s_i s_j}{2 \mu^4} + \ldots \right)  \exp\left[
-\frac{ \omega^2}{2 \Omega^2} - i \frac{\eta^2}{\mu^2} \bm{\omega} (\bm{r} - \bm{s}) - \frac{r^2 + s^2}{4  \hat{\rho}^2} \right] \hspace{35pt},
\label{remainder}
\end{eqnarray}
where summation over repeated indices is implied. The first terms in $R$ are the dipole and quadrupole components, whereas the quantity $\hat{\rho} = \left( \frac{1}{\rho^2} + \frac{2}{\mu^2}\right)^{-1/2} $ plays the role of a renormalized "mass".
 
We are interested in the structure of the phase diagram in the leading order in the small parameter of $\rho/\mu$. Since we are interested in the leading order, it appears from the equations that we may replace the renormalized $\hat{\rho}$ with the bare $\rho$.  Moreover,  within expressions (\ref{form3}) and (\ref{form4}) for $\hat{S}$ and $\hat{T}$, denominators of the form $\mu^2 + \rho^2 $ and $\mu^2 + 2 \rho^2$ can be approximated with $\mu^2$. 

In every integration where this kernel would play a role, variables representing relative LGN-cortex coordinates are confined by the arbor densities to a radius of order $\rho$ , and if we rely on the smallness of $\rho/\mu$ we can also rely on the smallness of the variables $r/\mu$, $s/\mu$ in their absolute values.  However these variables are also associate to angular directions which can lead to annihilating whole terms of an operator, no matter how large $\rho/\mu$ , when integrated over orthogonal angular component. Thus the smallness of $r/\mu$, $s/\mu$ cannot be used to discard 
(\ref{remainder}) by comparison with (\ref{L0_again}). 

The approximation we will purse is to discard all but the dipole term in $R$ (first term in expression \ref{remainder}). This is indeed the simplest restriction of Hilbert space that allows to explore whether, anywhere in the phase diagram, the system breaks out of circular symmetry. Doing so trasforms Eq.~(\ref{long_range_Lp}) into 
\bea
\nonumber 
L^p(\bm{r,s; \omega}) = e^{  - \frac{\omega^2}{2 \Omega^2}}   \bigg[ a^{*}_c(\bm{r}) a_c(\bm{s}) + \bm{A}^*(\bm{r}) \bm{A}(\bm{s}) + q^2 a_1^*(\bm{r}) a_1(\bm{s}), 
\\ 
- q \bigg(a^*_1(\bm{r}) a_d(\bm{s}) + 
a^*_d(\bm{r}) a_1(\bm{s}) \bigg) \bigg],
\label{reduced_rank}
\eea 
where $c=( \eta/\mu)^2$, $d = \frac{\eta^2+\rho^2}{\mu^2}$,  $q = \exp \left( - \frac{\zeta^4 \rho^2 \omega^2}{2 \mu^4} \right)$, and we have defined the functions
\be 
a_v(r) = \exp\left( i v\omega r_x - \frac{r^2}{4 \rho^2} \right) \ \ \ \ \ \bm{A}(\bm{r}) = \frac{\bm{r}}{\mu} \exp\left[ - \frac{r^2}{4 \rho^2} + i \frac{\eta^2}{\mu^2} \omega r_x \right], 
\ee 

where the index $v$ takes  the values $1$, $c$, and $d$. 

We now treat the $\eta \gg \rho$ and $\zeta \gg \rho$ cases separately, even if these assumptions will lead to similar results. 
  
\subsubsection{The regime $\eta \gg \rho$} 

If $\eta \gg \rho$, we have $d \sim c$, so the operator Eq.~(\ref{long_range_Lp}) becomes
\bea  \nonumber
L^p(\bm{r,s; \omega}) = e^{  - \frac{\omega^2}{2 \Omega^2}}   \bigg[ a^{*}_c(\bm{r}) a_c(\bm{s}) + \bm{A}^*(\bm{r}) \bm{A}(\bm{s}) + q^2 a_1^*(\bm{r}) a_1(\bm{s}) \\ 
- q \bigg(a^*_1(\bm{r}) a_c(\bm{s}) + 
a^*_c(\bm{r}) a_1(\bm{s}) \bigg) \bigg]
\label{eta_bigger_than_rho}
\eea 
 
Given one eigenfunction $\psi(r)$, let us now define the two unknowns $I_v = \int a_v (\bm{r}) \psi(\bm{r} ) \bm{d r}$ for $v=1,c$, and the third unknown $K= \int A_x(\bm{r}) \psi(\bm{r})  \bm{d r}$, and use the self-consistent assumption that $\int A_y(\bm{r}) \psi(\bm{r})  \bm{d r} =0$ (which is checked below in Sec.~\ref{section_long_range_analysis}). 
The eigenvalue equation for the operator of Eq.~(\ref{eta_bigger_than_rho})  becomes 
\begin{equation}
\lambda e^{  \frac{\omega^2}{2 \Omega^2}}   \psi(\bm{r}) = 
I_1 \left[ q^2 a_1^* (\bm{r})  - q a_c^* (\bm{r}) \right] + I_c \left[- q a_1^* (\bm{r})  + a_c^*(\bm{r}) \right] + K A_x^*(\bm{r}),
\label{eigenv_first}
\end{equation}
and computing the three unknown integrals from Eq.~(\ref{eigenv_first}) itself, one obtains: 
\begin{eqnarray}
\frac{\lambda}{2 \pi \rho^2}  e^{  \frac{\omega^2}{2 \Omega^2}}  I_1 &=& i \frac{\rho^2 \zeta^2}{\mu^3} \omega q  K; 
\\ 
\frac{\lambda}{2 \pi \rho^2}  e^{  \frac{\omega^2}{2 \Omega^2}}  I_c &=& 
\left(q^3 - q\right) I_1 + \left( 1 - q^2\right) I_c;
\\
\frac{\lambda}{2 \pi \rho^2}  e^{  \frac{\omega^2}{2 \Omega^2}}  K &=& 
-q^3 \frac{i \rho^2 \zeta^2}{\mu^3} \omega I_1
+ q^2 \frac{i \rho^2 \zeta^2}{\mu^3} \omega I_c 
+\frac{\rho^2}{\mu^2} K;
\end{eqnarray}
from which it follows that we can replace the infinite-dimensional operator of Eq.~(\ref{form1}) with the 3x3 matrix $ \hat{L} = 
2 \pi\rho^2 e^{ -\frac{ \omega^2}{2 \Omega^2}} \hat{M}$, where 
\begin{equation}
\label{3by3}
\hat{M} = \left( \begin{array}{ccc}
0 &0 & J q\\
q^3 - q  & 1- q^2 & 0 \\
-q^3 J & q^2 J & (\rho/\mu)^2 
\end{array} \right) 
\end{equation}
for $J = i \rho^2 \zeta^2 \omega /\mu^3$. 
  
This matrix has only two nonzero eigenvalues, both positive as we may expect from the discussion in Sec.~\ref{two_layers_positive_semid}. The larger one is 
\begin{eqnarray}
\nonumber 
\lambda &=& \pi \rho^2  e^{- \frac{\omega^2}{2 \Omega^2}} \Bigg( 1 + \rho^2 / \mu^2 - e^{- \omega^2 \zeta^4 \rho^2 /\mu^4 } \\ 
&& + 
\sqrt{\left(1 - \rho^2 / \mu^2 - e^{- \omega^2 \zeta^4 \rho^2 /\mu^4 } \right)^2 + \frac{4 \rho^4 \zeta^4 \omega^2}{\mu^6} e^{- 2 \omega^2 \zeta^4 \rho^2 /\mu^4 } } \Bigg),
\label{full_eigenvalue}
\end{eqnarray}

The corresponding eigenfunction is obtained from Eq.~(\ref{eigenv_first}) through the principal eigenvector of the matrix $\hat{M}$. 

This is found from Eq.~(\ref{3by3}) to be, before normalization,
 \begin{eqnarray}
I_1 &=& 2 J^2 q^3 \\ 
I_c &=& (1 - q^2)  \left[ 1 -s -q^2 + \sqrt{(1 -s - q^2)^2
-4 J^2 q^4} \right] 
\\ 
K &=& J q^2 \left[ 1 + s - q^2 + \sqrt{(1 -s - q^2)^2
-4 J^2 q^4}\right], 
\end{eqnarray}
with $s=(\rho/\mu )^2$. We now take the long-range limit as $s\rightarrow 0$ while keeping $q$ fixed, which yields 
\be  \left( \begin{array}{c} I_1, I_c, K \end{array} \right) \rightarrow 
\left( \begin{array}{c} 
0,1,0\end{array} \right), 
\ee 
hence the principal eigenfunction for the kernel Eq.~(\ref{eta_bigger_than_rho})
is found to be 
\bea
\nonumber 
\psi(\bm{r}) &=& \psi_1(\bm{r}; \omega_M \hat{\omega} ) \propto - q a_1^* (\bm{r}) 
+ a_c^*(\bm{r}) \\ 
&=& e^{  - \frac{r^2}{4 \rho^2}} \left[ e^{ - i \frac{\eta^2}{\mu^2} \omega_M \hat{\omega}\bm{r}} - e^{ - i\omega_M \hat{\omega} \bm{r} - \frac{1}{2} \left( \frac{\zeta^2 \rho \omega_M}{\mu^2} \right)^2} \right],
\label{full_eigenf}
\eea	
where $\hat{\omega}$ is an arbitrary unit vector. 
  
\subsubsection{The regime $\zeta \gg \rho$}

For $\zeta \gg \rho$, the eigenvalue equation of (\ref{reduced_rank}) can be written as
\begin{equation}
\label{eig_integr}
\lambda e^{  \frac{\omega^2}{2 \Omega^2}}  \psi(\bm{r}) = 
I_1 \bigg( q^2 a_1^*(\bm{r}) - q a^{*}_d (\bm{r})\bigg) + I_c a_c^*(\bm{r}) - I_d q a_1^*(\bm{r}) + K A_x^{*}(\bm{r}),
\end{equation}
where we defined the three unknown quantities $I_v = \int a_v (\bm{r}) \psi(\bm{r} ) \bm{d r}$ (for $v=1,c,d$) and the fourth unknown $K= \int A_x(\bm{r}) \psi(\bm{r})  \bm{d r}$. Again, we are using the self-consistent assumption that $\int A_y(\bm{r}) \psi(\bm{r})  \bm{d r} =0$, which will be duly checked in Sec.~\ref{section_long_range_analysis}. 
 
Define $J_x= \exp\left( - \frac{\rho^2 x^2 \omega^2}{2}\right)$, so that 
\begin{eqnarray}
q &=& J_{1-c};
\\
\int a_{\alpha}(\bm{r})
a_{\beta}^*(\bm{r}) d\bm{r} &=& 2 \pi \rho^2 J_{\alpha- \beta};
\\
 \int a_\beta(\bm{r}  ) A_x^* (\bm{r}) \bm{d r} &=& (2 \pi \rho^2)  i \frac{ \rho^2 \omega }{\mu} (\beta -c) J_{\beta -c};
 \end{eqnarray}
while $\int d\bm{r} A_x(\bm{r}) A^*_x(\bm{r}) = 2 \pi \rho^4/\mu^2$. 

From Eq.~(\ref{eig_integr}), we obtain 
\begin{eqnarray}
\frac{\lambda}{2 \pi \rho^2}  e^{\frac{\omega^2}{2 \Omega^2}} I_1&=&
\left(  q^2 - q J_{1-d} \right) I_1 + q I_c - q I_d + i \frac{\rho^2 \omega}{\mu} (1 -c) q K; 
\\ 
\frac{\lambda}{2 \pi \rho^2}  e^{\frac{\omega^2}{2 \Omega^2}} I_c&=&
(q^3 - q J_{d-c}) I_1 + I_c  -q^2 I_d;
 \\
\frac{\lambda}{2 \pi \rho^2}  e^{\frac{\omega^2}{2 \Omega^2}} 	I_d &=&
\left( q^2 J_{1-d} - q \right) I_1 + J_{d-c} -q J_{1-d} I_d + \frac{ i \omega \rho^4}{\mu^3} q 
K;
\\ 
\frac{\lambda}{2 \pi \rho^2}  e^{\frac{\omega^2}{2 \Omega^2}} K&=& 
\left( - i \frac{\omega \zeta^2 \rho^2 q^3}{\mu^3} + i q \frac{\omega \rho^4}{\mu^3} J_{d-c} \right) I_1 + i \frac{\omega \zeta^2\rho^2}{\mu^3} q^2 I_d + \frac{\rho^2}{\mu^2} K; 
\end{eqnarray}
from which it follows that, in this limit, we can replace our infinite-dimensional operator with the 4$\times$4  matrix
$ \hat{L} = 2 \pi \rho^2 e^{ -\frac{ \omega^2}{2 \Omega^2}} \hat{M}$, where 
\be 
 \hat{M} = \left( \begin{array}{cccc}
q^2 - q J_{1-d}&q& - q & \frac{ i \omega \zeta^2 \rho^2}{\mu^3} q \\
q^3 - q J_{d-c} & 1 & -q^2 & 0 \\
q^2 J_{1-d} - q & J_{d-c} & -q J_{1-d} & \frac{ i \omega \rho^4}{\mu^3} q 
\\ - i \frac{\omega \zeta^2 \rho^2 q^3}{\mu^3} + i q \frac{\omega \rho^4}{\mu^3} J_{d-c}
 & 0 &  i \frac{\omega \zeta^2\rho^2}{\mu^3} q^2 &  \frac{\rho^2}{\mu^2} \\
\end{array} \right). 
\ee  

Now, we have $d-c = \frac{\rho^2}{\mu^2} $ and $1-d= \frac{\zeta^2 - \rho^2}{\mu^2}$; since we are considering the regime where $\zeta \gg \rho$, we can write $1-d \sim \frac{\zeta^2}{\mu^2}$, so that $J_{1-d} \sim q$. Notice that we are making no assumption on the magnitude of $\eta$. The matrix thus simplifies to
\be 
 \hat{M} = \left( \begin{array}{cccc}
0 &q& - q & \frac{ i \omega \rho^2 \zeta^2}{\mu^3} q \\
q^3 - q J_{d-c} & 1 & -q^2 & 0 \\
q^3 - q & J_{d-c} & -q^2 & \frac{ i \omega \rho^4}{\mu^3} q 
\\ - i \frac{\omega \zeta^2 \rho^2 }{\mu^3} q^3+ i \frac{\omega \rho^4}{\mu^3} q J_{d-c} \ \ 
 & 0 &  i \frac{\omega \zeta^2 \rho^2}{\mu^3} q^2 &  \frac{\rho^2}{\mu^2} 
\end{array} \right). 
\label{the4by4matrix}
\ee  
  
Let us adopt one more self-consistent assumption, concerning the optimal wavenumber, which will be immediately verified once the optimal wavenumber is computed from the resulting eigenvalue. Namely, we assume  $ \omega \ll \frac{\mu}{\rho^2}$, so that we can write $J_{d-c} \sim 1$ and neglect the terms in $\omega \rho^4 /\mu^3$. The matrix Eq.~(\ref{the4by4matrix}) becomes 
\be 
\label{self-consistent-M}
 \hat{M} = \left( \begin{array}{cccc}
0 &q& - q & \frac{ i \omega \rho^2 \zeta^2}{\mu^3} q \\
q^3 - q & 1 & -q^2 & 0 \\
q^3 - q & 1& -q^2 & 0 
\\ - i \frac{\omega \zeta^2 \rho^2 }{\mu^3}  q^3\ \  & 0 \ \ &  i \frac{\omega \zeta^2 \rho^2}{\mu^3} q^2 &  \frac{\rho^2}{\mu^2} \\
\end{array} \right). 
\ee 

We have reduced an infinite dimensional problem to a four-dimensional problem, which we can solve exactly.  From Eq.~(\ref{self-consistent-M}), we see that 
\be 
\label{sec-det}
\det\left( \hat{M} - \lambda \right) = \lambda^2 \left[ \frac{\rho^2}{\mu^2}
- \frac{\omega^2 \rho^4 \zeta^4}{\mu^6} q^4 - \frac{\rho^2}{\mu^2} q^2 + \lambda \left(q^2 - 1 - \frac{\rho^2}{\mu^2} \right) + \lambda^2 \right] 
\ee 
and from Eq.~(\ref{sec-det}), it is found that the two non-null eigenvalues correspond to those of Eq.~(\ref{3by3}). Hence formula~(\ref{full_eigenvalue}) for the eigenvalue still holds true and, in particular, the optimal wavenumber will be the same in the two regimes. 

\subsection{The long-range limit: analysis of results} 
\label{section_long_range_analysis}

\subsubsection{Phase boundary and critical behavior} 
\label{long-range-phase-boundary} 

The system is in the T-phase if the wavenumber maximizing the principal eigenvalue is positive, while it is in either the R or N phase if that optimal wavenumber is null. In terms of the dimensionless variable $x= \omega^2 \zeta^4 \rho^2 /\mu^4$ (such that $q=e^{-x/2}$) we can write the principal eigenvalue~(\ref{full_eigenvalue}) as 
\bea
\label{2l_long_range_eigen} 
\lambda &=& \pi \rho^2 f(x),\\
\label{f}
f(x) &=& e^{-\alpha x/2}
\left( 1 +s - e^{-x} + \sqrt{ (1 - s - e^{-x})^2 + 4 s x e^{- 2 x} }\right),
\eea 
with $\alpha = \left(\frac{\mu \eta }{\zeta \rho}\right)^2$.

For $ \eta/\zeta$ of order one and $\mu \gg \rho$, $\alpha$ diverges, so the exponential prefactor in \ref{f} 
confines x to values of order $1/\alpha$. We can thus expand the expression in parenthesis in $x$ without any assumption on the magnitude of $s$, yielding   

$$ f(x) \sim 2 e^{- \alpha x /2 } (s + x ) $$  

with derivative 
$f'(x) \sim  e^{- \alpha x /2} ( 2 - \alpha s - \alpha x  ) $. Since this corresponds to a single maximum, the condition  for T-phase dominance is simply $f'(0) >0$ , i.e. 
\begin{equation} 
\label{RT_boundary_asymptotics}
\zeta > \zeta_c(\eta) = \sqrt{\frac{1}{2}} \ \eta. 
\end{equation}
whereas the wavenumber near the phase boundary  is given by 

$$ \omega \sim \frac{\mu}{\zeta^2} \sqrt{\frac{ 2 \zeta^2}{\eta^2} -1 } $$

\subsubsection{Form of the eigenfunction}

Separately pursuing as above the assumptions $\eta \gg \rho$ and $\zeta \gg \rho$ leads, as we saw, to the same eigenfunction~(\ref{full_eigenf}). This can be written as 
\begin{equation}
\label{LongRangeEigenf}
 \psi(\bm{r}) \propto e^{- \frac{r^2}{4 \rho^2} - i \frac{\eta^2}{\mu^2} \bm{\omega} \bm{r} } \left( 1 - e^{- i \frac{\zeta^2}{\mu^2}  \bm{\omega} \bm{r} - \frac{1}{2} \frac{\zeta^4}{\mu^4} \rho^2 \omega^2}\right). 
\end{equation}

\begin{figure}[htp]
\includegraphics[width=0.7\textwidth]{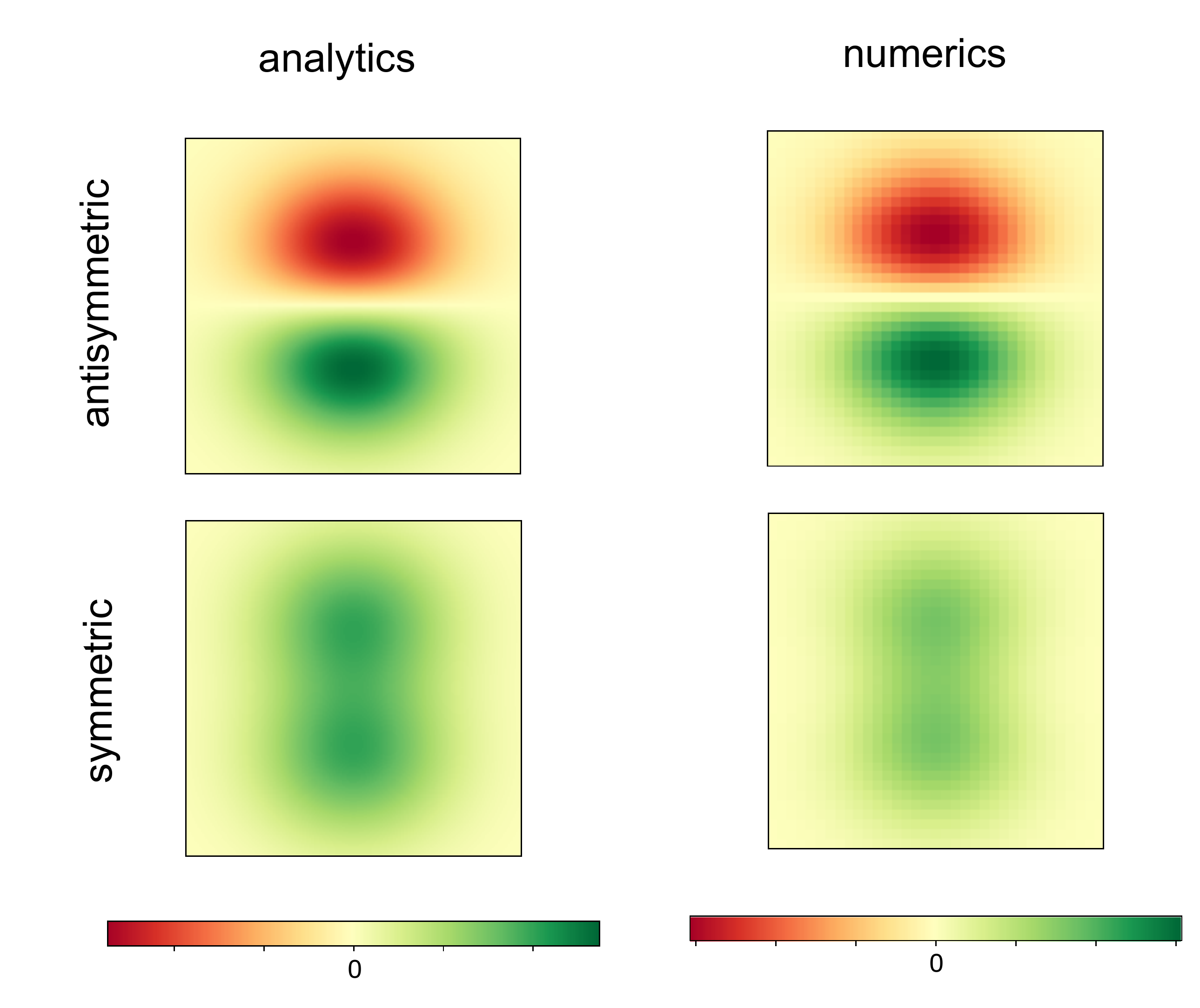}
\caption{
\small{\textbf{Receptive fields for the T-phase.}}
Plots of the symmetric and antisymmetric components of the approximation~\ref{psiT} to the Fourier-Transformed eigenfunction, compared to the result of numerically diagonalizing the full operator $\hat{L}^p$ and rescaling the principal eigenfunction by Eq.~\ref{t_to_s}. 
Receptive fields $\tilde{s}$ were rotated by the complex angle $\phi_0 =\arctan\left( - \int \Im \tilde{s} /\int \Re \tilde{s} \right)  $ so as to make the imaginary part  odd under inversion of the cortical modulation axis (the real part becomes symmetric as a consequence, see Sec.~\ref{sec:long-term-dyn}). 
The parameters used here are 
$\zeta=5\rho$ and $\eta=3\rho$, corresponding to a ground state at $\omega \approx 0.48 / \rho $. A  side of the grid has length equal to $5 \rho$, color scale ranges between min and values.}
\label{fig:T_check}
\end{figure}

Notice that the value of $\omega$ to be plugged into Eq.~(\ref{LongRangeEigenf}) is the value that maximizes the eigenvalue~(\ref{full_eigenvalue}). In regimes where the optimal wavenumber is null, we must take the $\omega \rightarrow 0$ limit in Eq.~(\ref{LongRangeEigenf}). Expanding the two complex exponentials to the first order in $\omega$ and keeping only the lowest order in the result yields the unnormalized eigenfunction
\be 
 \psi (\bm{r}) \propto r_x \exp\left( - \frac{r^2}{4 \rho^2} \right), 
 \ee 
equal to the orientation-selective eigenfunction we found for zero wavenumber region, that is, to an R-phase. (In those parts of the phase diagram, therefore, the homeostatic constraint is satisfied through the individual selectivity of cells, and does not need to be satisfied through variations over cortical space; that is why translation symmetry can be restored.)
 
With $\psi(\bm{r};\bm{\omega})$ given by Eq.~(\ref{LongRangeEigenf}), the eigenfunctions $\psi(\bm{x}, \bm{r}) = \psi(\bm{r};\bm{\omega}) e^{i \bm{\omega}\bm{x}}$ for wave vectors $\bm{\omega}$ and $-\bm{\omega}$ are degenerate and complex conjugates of each other. A real linear combination of the two is obtained by taking either the real or imaginary part. From this, via Eq.~\ref{t_to_s},  formula 
\ref{psiT} of the main text is obtained (see Fig.~\ref{fig:T_check}).

\subsubsection{Normally oriented eigenfunctions}
  
In order to obtain Eq.~(\ref{full_eigenf}),  we made at the very outset (below Eq.~\ref{eta_bigger_than_rho}) the self-consistent assumption $\int A^*_y(\bm{r}) \psi(\bm{r}) d\bm{R} = 0 $, which we used to write both Eq.~(\ref{eigenv_first}) and (\ref{eig_integr}). The subspace we have focused on was indeed orthogonal to $A_y$, and we found this subspace to be an asymptotic eigenspace of the system -- see Eq.~(\ref{LongRangeEigenf}).

Nonetheless, the same system may also possess eigenfunctions having a nonzero overlap with $A_y(\bm{r})$. Do these eigenfunctions correspond to a higher eigenvalue than those we calculated? 
 
The self-consistency of our initial assumption is straightforward to check. If we repeat the above by relaxing the assumption $\int A^*_y(\bm{r}) \psi(\bm{r}) d\bm{R} = 0 $, we have to diagonalize  a $5\times 5$ matrix instead of a $4\times 4$ one. However, this matrix is diagonal in its $A_y$-sector. The resulting extra eigenvalue is a strictly decreasing function of wavenumber, hence it must be computed at $\omega = 0$, where we find  $ \Lambda_y = 2 \pi \rho^4 /\mu^2$.  

Let us compare this eigenvalue with the eigenvalue of the cortically modulated solution $\Lambda_M$ which we found above. In the regions of the phase diagram where $\omega_M >0$, we have $\Lambda_y <  \Lambda_M$, hence the normally oriented solution is suppressed at long times. In the regions where optimal wavenumber is $\omega_M = 0 $, on the other hand, it can be seen that $\Lambda_y = \Lambda_M$ .
 
We thus find that, when the RF varies across the cortex, it tends to vary from negative to positive along the direction of cortical modulation, so that the orientation is orthogonal to that direction.  When it is uniform across the cortex, its direction becomes immaterial, hence we have degeneracy in the orientation of the RF.  
 
\section{The uncorrelated regime} 
\label{sec:uncorrelated}
 
\subsection{The operator with uncorrelated input ($\zeta \sim 0$)} 
\label{sec:uncorrelated_op}    
    
In order to infer the other main feature of the phase diagram, i.e. the existence of a triple point, we must focus on the uncorrelated limit  $\zeta \ll \min (\rho, \eta)$. In this limit, it follows from Eq.~(\ref{unconstrained_matr_el}) that the matrix element of $\hat{L}^p$ in real space takes the form 
\be 
\label{Lp-for-low-zeta}
L^p(\bm{x},\bm{r}; \bm{y}, \bm{s}) = 
\delta\left( \bm{x} - \bm{y} + \bm{r} -\bm{s}\right) L^c(\bm{r}, \bm{s}),
\ee 
where the operator $\hat{L}^c$  has the kernel
\bea 
\nonumber 
L^c (\bm{r}, \bm{s}) &=& \int d \bm{r}_1 d \bm{s}_1 \left[ \delta ( \bm{r} - \bm{r}_1) - \sqrt{A(\bm{r} ) A(\bm{r}_1)} \right] 
L^{0}(\bm{r}_1, \bm{s}_1) \\ 
&& \times 
\left[ \delta ( \bm{s_1} - \bm{s}) - \sqrt{A(\bm{s}_1) A(\bm{s})}\right], 
\label{LcExpression}
\eea
with 
\be
\label{L0_matrix_el}
L^{0}(\bm{r},\bm{s}; \eta) = \sqrt{A(r) A(s) } I(\bm{r},\bm{s})
\ee
where to write the last equality we have applied the delta function of equation \ref{Lp-for-low-zeta} to infer 
$I(\bm{x,y})=I(\bm{x-y})=
I(\bm{r-s})$.   

Implementing the delta functions in Eq.~(\ref{LcExpression}), we have
\begin{eqnarray}\nonumber
\frac{ L^c(\bm{r}, \bm{s}) }{ \sqrt{A(r)  A(s)} } = I  \left( \bm{r},\bm{s} \right) 
+
\int d\bm{r}_1 d \bm{s}_1 A(r_1) A(s_1) I  \left( \bm{r}_1 , \bm{s}_1\right) \\ -  \int d\bm{u} A(u) I  \left( \bm{r} , \bm{u} \right)
 -  \int d\bm{u} A(u) I  \left( \bm{u} , \bm{s} \right) 
\label{Lc_to_integrate}
\end{eqnarray}

We can also rewrite  Eq.~(\ref{LcExpression}) compactly as:
\begin{equation}
\label{Lc_operator}
\hat{L}^c = \hat{L}^{0} + | a_0 \rangle \langle a_0 | \hat{L 
}^{0} | a_0 \rangle \langle a_0 | - 2 \ \text{HP } \left[ \hat{L 
}^{0}| a_0 \rangle \langle a_0 |   \right],
\end{equation}
where "HP" is the Hermitian part of an operator.  
 
If we look for eigenfunctions of $\hat{L}^p$ in the form 
\be 
\label{Psi-form}
\Psi(\bm{x}, \bm{r}) \equiv \langle \bm{x}, \bm{r} | \Psi \rangle = \psi(\bm{r}) e^{-i \bm{\omega} (\bm{x + r})},
\ee
the characteristic equation $\Lambda \Psi = \hat{L}^p \Psi$ reduces to 
$  \Lambda \psi(\bm{r}) 
 = \int \bm{d s} \ L^c(\bm{r}, \bm{s} ) \psi(\bm{s}) $, which means that $\psi(\bm{r})$ is the corresponding eigenfunction of $\hat{L}^c$ and the eigenvalue is independent of the cortical wavenumber. Hence, in the limit of uncorrelated inputs there is complete degeneracy in the wavenumber.  

Due to this degeneracy the principal eigenfunction could be calculated by focusing solely on the zero-wavenumber sector. The results of section \ref{sec:uniform}  apply and can be used to  compute $\psi(\bm{r})$ which is then   replaced for $\psi(\bm{r})$ in Eq.~(\ref{Psi-form}), yielding the principal eigenfunction for all wavenumbers at $\zeta=0$.  

Thus, we find  that if the input is completely uncorrelated ($\zeta =0$) , the principal eigenfunction of the operator is orientation selective if $\eta  > \Theta_c \rho $ and non-selective if $\eta < \Theta_c \rho $, with the critical ratio $\Theta_c $ bounded from below by the value $\theta_c$ of Eq.~(\ref{theta_c}). 
   
For $\zeta=0$ and $\eta <\theta_c \rho $,
in the variational approximation of Eq.~(\ref{moveable_node}),
the principal eigenfunction is 
 \be 
  \Psi_0(\bm{x}, \bm{r}) \propto (R^2 - r^2) \exp\left(  -\frac{r^2}{2 \sigma^2} + 
i \omega (x+r_x)  
\right), 
\ee
where $\sigma$ is the value of $\gamma$ as given by Eq.~(\ref{redefinition}) evaluated at $\zeta=0$.

The nodal radius $R \propto \sqrt{\eta \rho}$  is as calculated in Sec.~\ref{sec:uniform}
and the eigenvalue is $\lambda \sim \frac{2 \pi \eta^3 }{\rho} \left( \rho - (5 - \sqrt{10}) \eta\right) $. While the eigenfunction depends parametrically on the wavenumber $\omega$, the eigenvalue is entirely degenerate in it, as follows from the divergence of the cutoff wavenumber $\Omega$ in Eq.~(\ref{mu-and-omega}). 
 
For $\zeta=0$ and $\eta > \Theta_c \rho $, the exact principal eigenfunction as 
per Eqs.~(\ref{0_1_eigenf},\ref{1_0_eigenf}) is given by
\be  \Psi(\bm{x}, \bm{r})= \left(k_+e^{+i\phi} + k_- e^{-i\phi}\right) \exp\left[ - i \bm{\omega (x+r)  } - \frac{r^2}{2 \sigma^2} \right], 
\ee
for any vector $\bm{\omega}$ and arbitrary coefficients $k_+$ and $k_-$.  Again we refer to the longitudinal and orthogonal combinations 
\begin{equation}
\left( \begin{array}{c}
\Psi^{x}(\bm{x}, \bm{r}) 
\\
\Psi^{y}(\bm{x}, \bm{r}) \end{array} \right) = 
\left( \begin{array}{c}
r_x \\
r_y
\end{array} \right)  \exp\left(  -\frac{r_x^2 + r_y^2}{2 \sigma^2} + 
i \bm{\omega (x+r)} \right),  
\end{equation}
which are the instances of the 2p waves $\Phi_{0,1}$ and $\Phi_{1,0}$ corresponding to uncorrelated input. 

These two eigenfunctions share, as per Eq.~(\ref{Linsker_spectrum}), 
the exact eigenvalue
\be 
\Lambda^{x,y} =   2 \pi \eta^2 \left( 1 + \frac{\eta^2}{2 \rho^2} + \frac{\eta}{\rho} \sqrt{1 + \frac{\eta^2}{4 \rho^2} } 
\right)^{-2};
\ee
which is independent on the wavenumber. Thus, the x-y degeneracy we have for $\eta > \Theta_c \rho$ adds up to the overall degeneracy in the cortical wavenumber that exists for any value of $\eta$. 

\subsection{Perturbative input correlations} 
\label{sec:perturbative}  
 
We have shown that the point $P_0 = (\zeta_0 = 0, \eta_0 = \Theta_c \rho)$  where the R and N phase meet is a point of non-analyticity for the principal eigenvalue regarded as a function of the parameters, and thus belongs to a phase boundary. Moreover, this phase boundary cannot stop there, because it is a boundary between two phases that have different symmetries -- one that displays orientation selectivity and one that does not.  How is this phase boundary continued for $\zeta >0$? Will it curve up or down in the $(\zeta,\eta)$ space?
  
Since we possess the exact solution for $\zeta=0, \eta > \Theta_c \rho$, perturbation theory is an ideal tool to address this question. We will build a perturbation theory in the small parameter  $\zeta/\eta$. As we saw in Sec.~\ref{sec:uncorrelated_op}, our starting point for perturbation theory is a highly degenerate set of eigenfunctions, mainly due to the degeneracy in the wavenumber. But since the full operator for $\zeta >0$ commutes with cortical translations,  different translational eigenstates are not coupled by the perturbation, and nondegenerate perturbation theory with respect to wavenumbers may be applied.
   
The theory will prove the following three facts: 
 
(1) the $\omega$ degeneracy is removed by an infinitesimal $\zeta>0$ for any $\eta$, and this happens in such a way that $\omega=0$ is always the principal eigenstate; 

(2) the phase boundary starting at the point $P_0=(0,\Theta_c \rho)$ has a flat slope at that point in the $\zeta/\eta$ plane;
 
(3) the $xy$ degeneracy of the p-wave eigenfunctions survives at finite $\zeta$.

\subsection{Optimal wavenumber for oriented eigenfunctions: transverse orientations}
\label{sec:oriented-transverse}

We have mentioned that the perturbation does not couple degenerate wavenumbers. The same is true with the additional degeneracy in the orientation of selectivity, and it is possible to study the two 2p eigenfunctions separately because the full operator $\hat{L}^p$ does not couple them for any value of the parameters.  Indeed, we have it by symmetry that $ \langle \Psi^x | \hat{L}^p | \Psi^y \rangle =0$ for any $\zeta$ and $\omega$. This means that we can study the effect of a small but finite $\zeta$ separately on the two eigenfunctions (applying nondegenerate perturbation theory). 
 
We begin with the y-oriented wave (transverse orientation). While $\Psi^y$ is an eigenfunction of $\hat{L}$ only for $\zeta =0$, it can be checked that its generalization 
\be 
\chi^y_{0,1} (\bm{r}) = \frac{r_y}{\sqrt{\pi} \sigma^2} \exp\left(  - \frac{r^2}{2 \sigma^2} \right) e^{- i \frac{\eta^2}{\mu^2} \bm{\omega r}}, 
\ee 
is an exact eigenfunction of the full operator $\hat{L}^p$ over the whole phase diagram. Indeed, it is an exact eigenfunction of $\hat{L}$ and, being orthogonal to the constraint ket $|a_{\omega}\rangle
 $, it belongs to the null space of the constraint operators $\hat{S}$ and $\hat{T}$: 
 
 \be 
 \langle \chi^y_{0,1} | \hat{S} | \chi^y_{0,1} \rangle = \langle \chi^y_{0,1} | \hat{T} | \chi^y_{0,1} \rangle = 0. 
 \ee 

The corresponding eigenvalue of $\hat{L}^p$ is given, for every point in $(\zeta,\eta)$ space, by 
\begin{eqnarray}
\label{y_eig}
\Lambda^y = \frac{2 \pi \mu^2 }{\beta^2} \exp\left( -\frac{ \omega^2}{2 \Omega^2} \right), 
\end{eqnarray}
with $\beta$ and $\Omega$ defined according to Eq.~(\ref{redefinition}) and (\ref{mu-and-omega}) respectively. 
 
For $\zeta =0$, as we knew, this eigenvalue is independent on the wavenumber. However, for any $\zeta > 0$, Eq.~(\ref{y_eig}) describes an eigenvalue that decreases monotonically with the wavenumber, hence the degeneracy is removed. We can conclude that, in the limit of small $\zeta$, the principal y-oriented eigenfunction is uniform over the cortex, i.e. translation symmetry is not broken. 

\subsection{Optimal wavenumber for oriented eigenfunctions: longitudinal orientations} 
\label{sec:oriented-longitudinal}

We now turn to considering the x-oriented function $\Psi^x$. It can be checked that $\Psi^x$ is orthogonal to the constraint state $|a_{\omega}\rangle$, which entails
\be 
\label{orthogonality}
 \langle \Psi^x | \hat{S} | \Psi^x \rangle = \langle \Psi^x | \hat{T} | \Psi^x \rangle = 0, 
 \ee 
This holds true for any value of $\zeta$. However $\Psi^x$ is an eigenstate only for $\zeta=0$ and, differently from the case of $\Psi^y$ seen above, it is not straightforward to build a generalization of 
 $\Psi^x $ that will be an eigenstate of $\hat{L}^p$ at any point in parameter space. 
   
Therefore, we will restrict here our attention to sufficiently small non-zero values of $\zeta$ and will build a perturbation theory in the parameter $\epsilon = \zeta^2/\eta^2$. Hence we write the operator $\hat{L}^p$ as $\hat{L}^p = \hat{L}^p(\epsilon=0)  + \hat{\Delta} + O(\epsilon^2)$, where $\hat{\Delta}$ includes the first order in $\epsilon$, and we will treat $\hat{\Delta}$ as a perturbation.  
In the shift operator $\hat{\Delta} = \Delta \hat{L} + \Delta \hat{S} + \Delta \hat{T} + \Delta \hat{T}^\dagger$, because of Eq.~(\ref{orthogonality}), we only have to compute the $\hat{L}$-term. 
 
We begin by expanding to the first order in $\epsilon$ Eq.~(\ref{form2}), which yields 
\be
\label{DeltaL}
\Delta L (\bm{r}, \bm{s}; \omega) = \epsilon \left[ - \frac{\omega^2 \eta^2}{2} + i \omega (r_x - s_x)+ \frac{(\bm{r}- \bm{s})^2}{2 \eta^2}  \right] e^{ - i \omega (r_x - s_x) - \frac{r^2 + s^2}{4 \rho^2} - \frac{(\bm{r - s})^2}{2 \eta^2} }. 
\ee
We only keep the terms that have a nonvanishing expectation value in $\Psi^x$; in particular, we neglect terms that change sign if we swap the two variables $r_x $ and $ s_x$, because the integral would be zero. In addition, we may ignore terms whose expectation value in $\Psi^x$ (i.e. whose contribution to $\langle \Psi^x | \Delta \hat{L} | \Psi^x\rangle$) will bear no  dependence on the wavenumber. After some algebra this leaves a single first-order term in Eq.~(\ref{DeltaL}) that obeys all these requirements, namely: 
\begin{eqnarray}
\Delta L (\bm{r}, \bm{s}; \omega) \sim  - \frac{\omega^2 \zeta^2}{2} e^{ - i \omega (r_x - s_x) - \frac{r^2 + s^2}{4 \rho^2} - \frac{(\bm{r - s})^2}{2 \eta^2}}. 
\end{eqnarray}
 
The corresponding expectation value is 
\be 
\label{deltadelta}
 \Delta =\langle \Psi^x | \Delta \hat{L}| \Psi^x \rangle = - \frac{\omega^2 \zeta^2}{2} \Lambda_{2 p}(\zeta=0), 
 \ee
a negative shift in the eigenvalue that is minimized by setting $\omega=0$. 

We have thus proven that, for sufficiently small $\zeta$ and given $\eta$, the principal eigenstate is always cortically uniform ($\omega=0$),  as long as the principal eigenstate for the given $\eta$ and $\zeta=0$ is an R-phase. This entails that in the limit $\zeta \rightarrow 0$, the principal eigenstate of the system has a zero wavenumber for any $\eta > \Theta_c \rho$. Hence, the slope of the phase boundary at $(0, \Theta_c \rho)$ cannot be positive.  
  
While the degeneracy in wavenumber has been removed by first-order perturbation theory, the degeneracy between the $x$ and $y$ orientations has not been removed, as seen by comparing Eq.~(\ref{y_eig})
 and Eq.~(\ref{deltadelta}) to the second order in $\omega$ and using $\Omega \sim 1/\zeta$. 
 
\subsection{Optimal wavenumber for non-oriented eigenfunctions} 
 
The s-wave (that is, non-oriented) eigenfunctions of $\hat{L}^p$ are also degenerate in the cortical wavenumber for $\zeta=0$. To see which wavenumber effectively prevails, we must build a perturbation theory in $\epsilon = \zeta^2/\eta^2$ starting from the (unknown) principal eigenstate of the zero-wavenumber operator, which we call $| s \rangle$  because of its being an s-mode. 

From the discussion of Sec.~\ref{sec:uncorrelated} (see Eq.~\ref{Psi-form}), we know that the the principal eigenstate at $\zeta=0$, when an s-wave, must have the form $\langle \omega, \bm{r} | s \rangle = \psi_s (r) e^{- i \omega r_x}$. We can thus use Eq.~(\ref{proj_form}) to write the level shift as 
\be 
\label{sshift}
\Delta \hat{L}^p (\omega) = \langle s | \left(1 - | a_\omega \rangle \langle a_\omega | \right) 
\Delta \hat{L}_\omega \left( 1 - | a_\omega \rangle \langle a_\omega |\right) | s \rangle 
\ee
where the matrix elements of $\Delta \hat{L}_\omega $ have the form given in Eq.~(\ref{DeltaL}). 
 
We notice now that factors of the type $e^{- i \omega r_x}$ will cancel in the integrands of all scalar products that appear in Eq.~(\ref{sshift}). As a consequence, the third term in the square brackets of Eq.~(\ref{DeltaL}) may be ignored, as it adds no dependence on the cortical wavenumber. 

The first term in the square brackets of Eq.~(\ref{DeltaL}), on the other hand, yields the level shift
\be 
\label{DeltaL1}
\Delta \hat{L}_e^{(1)} = - \frac{ \omega^2 \zeta^2}{2} \langle s | \hat{L}^p (\zeta=0) | s \rangle = - \frac{ \omega^2 \zeta^2}{2} \int \psi_s (\bm{r}) L^c (\bm{r}, \bm{s})\psi_s(\bm{s}) \bm{d r}  = - \frac{ \omega^2 \zeta^2 \Lambda_s }{2} 
\ee 
where to write the last equality we have used the fact that $\psi_s$ is, by definition, an eigenfunction of $\hat{L}^c$ with a positive eigenvalue $\Lambda_s$. The resulting shift is a monotonically decreasing function of the wavenumber. 
   
The only remaining term is the second one in the square brackets of Eq.~(\ref{DeltaL}), namely
\be 
\label{deltaL2}
\Delta L^{(2)} (\bm{r}, \bm{s}; \omega) = i \epsilon \omega (r_x - s_x) e^{ - i \omega (r_x - s_x) - \frac{r^2 + s^2}{4 \rho^2} - \frac{(\bm{r - s})^2}{2 \eta^2} },
\ee
which also yields a level shift of the form  
\be 
\label{shiftL2}
\Delta \hat{L}_e^{(2)} = \langle s | \Delta \hat{L}^{(2)}  | s \rangle  + | \langle s | a_\omega \rangle |^2 \langle a_\omega | \Delta \hat{L}_\omega | a_\omega \rangle - 2 \ \Re \left[ \langle s | a_\omega \rangle \langle a_\omega | \Delta \hat{L}^{(2)}  | s \rangle\right].
\ee
The ensuing integrals are quickly estimated by symmetry considerations. The factors $e^{-i \omega (r_x - s_x) }$ in Eq.~(\ref{deltaL2}) cancel everywhere in Eq.~(\ref{shiftL2}). While only the real part of these imaginary exponentials would contribute, the resulting matrix element is effectively antisymmetric in the swapping of the $\bm{r}$ and $\bm{s}$ coordinates. This makes the first two terms in Eq.~(\ref{shiftL2}) vanish by symmetry; since the final square bracket is purely imaginary, the third term is also zero. 

Hence the full eigenvalue shift is given by Eq.~(\ref{DeltaL1}) and decreases monotonically as a function of the wavenumber. We conclude that the wavenumber degeneracy at $\zeta=0$ is removed even by an infinitesimal range of presynaptic correlations, and the uniform cortical mode is favored for any $\eta$.  

Building upon this, we conclude that, to the lowest order in $\zeta/\eta$, the principal eigenvalues of the s- and p-modes are unchanged from those at $\zeta=0$, and therefore the phase boundary starting from $P_0 = (0, \Theta_c \rho)$  has a flat slope at that point (Fig.~\ref{fig:chart_tp}).
 
\section{The triple point}

\begin{figure}[htp]
\includegraphics[width=0.6\textwidth]{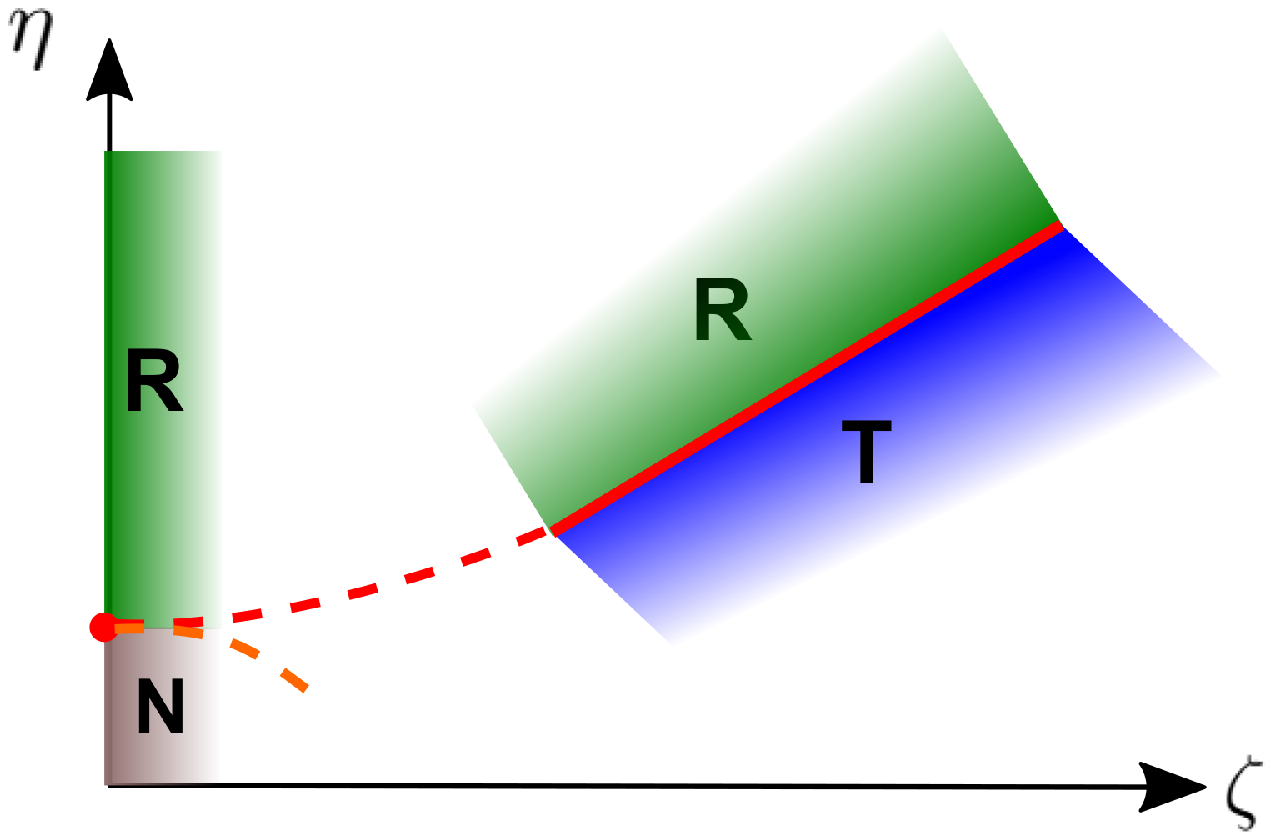}
\centering
\caption{\small{\textbf{Patching together of phase boundaries.}}
For uncorrelated inputs (Sec.~\ref{sec:uncorrelated}), the $\eta$-axis contains a transition point where rotation symmetry is broken (red dot). The perturbation theory for short-range input correlations (Sec.~\ref{sec:perturbative}) shows that this transition point is continued by a flat phase boundary. The  asymptotic rank reduction used for the long-range limit (Sec.~\ref{sec:long_range}), revealed an RT boundary far away from the origin. We also know from Sec.~\ref{sec:uniform_boundary} that the N phase is forbidden outside a quarter-circle containing the red dot on its contour, and the R phase is forbidden inside it. This leads to predicting an N-T boundary (orange dashed line) contained within the quarter-circle, and an R-T boundary stretching from the red dot into the long-range regime. The red dot is a triple point of the system.} 
\label{fig:chart_tp}
\end{figure}

We pointed out the existence of a point $P$ on the $\zeta=0$ axis where the $N$-phase transitions into the $R$-phase, and showed that this phase boundary continues parallel to the $\zeta$-axis for perturbatively small values of $\zeta$. This must be matched with what was shown about the long-range limit, the existence of a linear phase boundary between the R and T phase. These boundary lines cannot terminate, but can only continue into each other, the reason being that beyond the termination point of a phase boundary two different symmetries would have to merge. The simplest diagram  adhering to this requirement is one where the R-region extending above the TR boundary for long ranges connects, at short ranges, to the R-region that extends above the NR boundary. The missing stretch of phase boundary is sketched as a  dashed red curve in  Fig.~\ref{fig:chart_tp}.

The immediate consequence of this scenario is that an extra boundary, located at values of $\eta$ below the lower limit of the R-phase, must separate the short-range N-region from the T-region that begins for  longer ranges (orange dashed curve of Fig.~\ref{fig:chart_tp}). In principle one could expect this second boundary to start from any point along the lower contour of the R-phase, i.e. at an arbitrary value of $\zeta$. However, it must be remembered the discontinuity between the R and N eigenfunctions is only necessary for $\zeta=0$, where projective fields from LGN are decoupled from each other and there is degeneracy in the wave number. For $\zeta>0$, a suitable path through the T-phase can always bridge the N and R eigenfunctions continuously. Variational reasoning is sufficient to infer that, where such transition exists, it is advantageous over a sharp transition. 

It follows that we expect the T region to taper all the way to point $P_0$, which consequently is a triple point of the system.  

\end{document}